\documentclass[reprint,superscriptaddress,aps,prl,twocolumn,longbibliography]{revtex4-2}
\usepackage{makeidx}

\usepackage[utf8]{inputenc}
\usepackage[T1]{fontenc}
\usepackage{graphicx}
\usepackage{mathrsfs}
\usepackage{braket}
\usepackage[subrefformat=parens,labelformat=parens,caption=false]{subfig}
\usepackage{amsfonts}
\usepackage{amssymb}
\usepackage{amsthm}
\usepackage{amsmath}
\usepackage{mleftright}
\usepackage{dcolumn}
\usepackage{color}
\usepackage[dvipsnames]{xcolor}
\usepackage{verbatim}
\usepackage{cancel}
\usepackage{multirow}
\usepackage[breaklinks,
            colorlinks,
            urlcolor=blue,
            linkcolor=blue,
            anchorcolor=blue,
            citecolor=blue]{hyperref}

\usepackage{cleveref}
           
\usepackage{soul}
\newcommand\norm[1]{\left\lVert#1\right\rVert}

\newcommand{\Eset}[1]{\underset{#1}{\mathbb{E}}}

\newcommand{\parens}[1]{\left(#1\right)}
\newcommand{\sparens}[1]{\left[ #1 \right]}

\newcommand{\fsre}[0]{\widetilde M}
\newcommand{\tr}[0]{\text{Tr}}
\newcommand{\dd}[1]{\text{d}#1}
\newcommand{\comm}[1]{}
\DeclareMathOperator{\E}{\mathbb{E}}

\newcommand{\ps}[0]{\Xi}
\newcommand{\ipr}[0]{\zeta}

\newcommand{\bigO}[1]{\mathcal{O}\mleft(#1\mright)}

\DeclareMathOperator{\Tr}{Tr}

\definecolor{THc}{rgb}{0.9,0.3,0.2}

\newcommand{\prlsection}[1]{{\em {#1}.---}}

\newcommand{\SM}{SM}

\begin{document}

\title{Universal equilibrium magic in quantum many-body systems}

\author{Soumyadeep Sarma}
\affiliation{Department of Physics, Indian Institute of Science, Bangalore 560012, India}
\author{Tobias Haug}
\affiliation{Quantum Research Centre, Technology Innovation Institute, Abu Dhabi, UAE}
\author{John Preskill}
\affiliation{Institute for Quantum Information and Matter, California Institute of Technology, Pasadena, CA 91125, USA}
\author{Wai-Keong Mok}
\affiliation{Institute for Quantum Information and Matter, California Institute of Technology, Pasadena, CA 91125, USA}

\begin{abstract}
Thermalization conventionally describes local properties of isolated many-body systems at equilibrium. Magic, or nonstabilizerness -- the resource enabling universal quantum computation -- is by contrast encoded in the global structure of the many-body wavefunction. We show that, despite its global nature, the magic of equilibrium pure states of chaotic many-body systems, including late-time evolved states and energy eigenstates, is universal: it is captured by the thermal Scrooge ensemble, the minimally informative ensemble of pure states consistent with the Gibbs state at the same effective temperature. Therefore, for systems with no conserved quantities other than the total energy, equilibrium magic is a function of temperature alone, independent of the initial state and other microscopic features of the equilibrium state. This yields concrete universal predictions for the stabilizer R\'enyi entropies (SREs). At infinite temperature, the SRE is set by Haar-like fluctuations of the Pauli spectrum, while at finite temperature energy conservation induces a volume-law thermodynamic correction controlled by the thermal Pauli spectrum. We support these predictions with analytical arguments and extensive numerical simulations. We further show that chaotic many-body systems at high temperatures possess long-range magic and entanglement that cannot be removed by finite-depth local quantum circuits. Our results establish magic as a thermodynamic property of chaotic many-body systems and suggest that Scrooge ensembles may provide a unified framework for quantum many-body resources.

\end{abstract}
\maketitle

 \let\oldaddcontentsline\addcontentsline\renewcommand{\addcontentsline}[3]{}

\prlsection{Introduction} 
Isolated quantum many-body systems can exhibit remarkably simple behavior at long times~\cite{rodriguez2024eeinfinitetemp,langlett2025eefinitetemp,rigol2008thermalization}. Although their microscopic dynamics is unitary and reversible, local observables in generic non-integrable systems relax to values described by equilibrium statistical mechanics~\cite{deutsch1991quantum}. This emergence of thermodynamics from quantum dynamics is commonly understood through quantum typicality and the eigenstate thermalization hypothesis (ETH), according to which highly excited eigenstates behave locally as thermal states~\cite{deutsch1991quantum,srednicki1994chaos,srednicki1999approach,dalessio2016quantum,dymarsky2018subsystem,deutsch2018eigenstate}. A central lesson of this framework is that entanglement, despite being a property of pure quantum states, acquires a thermodynamic structure in chaotic many-body systems: its behavior is governed by equilibrium entropy rather than by microscopic details~\cite{vidmar2017entanglement,lu2019renyi,garrison2018does,murthy2019structure,bianchi2022volume,rodriguez2024eeinfinitetemp,langlett2025eefinitetemp}.

Much less is known about whether an analogous principle governs \emph{magic}, or nonstabilizerness~\cite{Veitch2014ResourceTheoryStabilizer}. 
Magic is a necessary ingredient for quantum advantage, which combined with classically efficient Clifford operations allows for  universal quantum computation~\cite{bravyi2005universal,bravyi2016trading,gottesman1997stabilizer}.

Recent work has shown that highly excited chaotic eigenstates near infinite temperature have magic close to that of Haar-random states~\cite{turkeshi2025fSRE}. However, generic many-body states at finite energy densities correspond to finite effective temperatures: energy conservation constrains their local properties, and it is not a priori clear how such thermodynamic constraints can shape magic, which is a highly non-local property that depends on the global structure of the quantum state.

In this Letter, we show that magic in chaotic quantum many-body systems is governed by a universal thermodynamic principle. 
We discover that the distribution of Pauli expectation values (i.e., the Pauli spectrum~\cite{beverland2020lower}) of equilibrium pure states, including highly excited eigenstates and late-time states evolved from simple product states, is well described by that of the thermal Scrooge ensemble.
The Scrooge ensemble is the maximally informationally stingy ensemble compatible with a specified density matrix~\cite{jozsa1994scrooge,mark2024deeptherm,mok2026universalityscrooge}: it generalizes Haar-random states by incorporating physical constraints, such as energy conservation, while minimizing the accessible information of the ensemble. 
For a chaotic Hamiltonian with no conserved quantities beyond energy, the relevant density matrix is the Gibbs state $\sigma_\beta \propto e^{-\beta H}$, with inverse temperature $\beta$ fixed by the energy density of the equilibrium pure state. 
We find that this single thermodynamic input determines the leading behavior of many-body magic, and we support this claim with extensive numerical simulations.

\begin{figure}
    \centering
    \includegraphics[width=0.9\columnwidth]{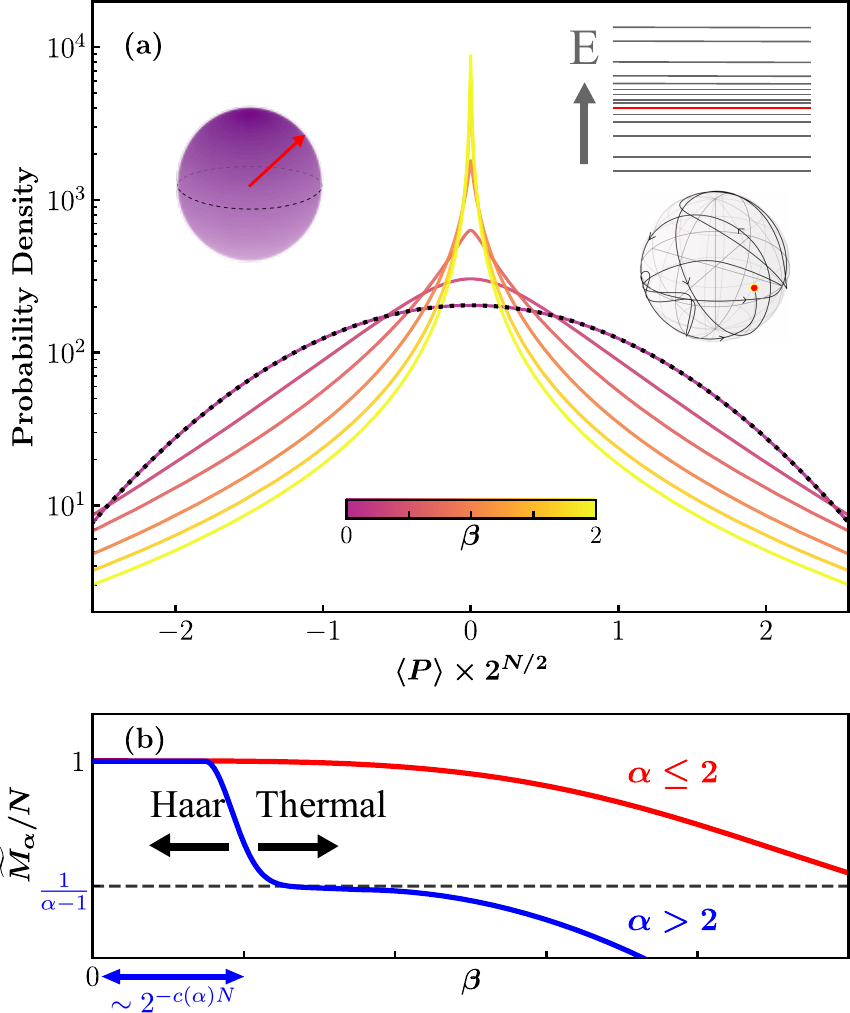}
    \caption{Schematic of the main results. (a) Pauli spectrum distribution for a typical state drawn from the thermal Scrooge ensemble of a chaotic local $N$-qubit Hamiltonian, shown for inverse temperatures $\beta \in [0,2]$. At $\beta = 0$, this reduces to Haar-random states whose Pauli spectrum is close to the Gaussian distribution $\mathcal{N}(0,2^{-N})$~\cite{turkeshi2025fSRE} (black dashed line). For $\beta = \bigO{1}$, the same Scrooge description captures the Pauli spectrum, and hence the magic, of equilibrium pure states such as energy eigenstates and late-time evolved states. (b) Magic density $\widetilde{M}_\alpha/N$ versus $\beta$. For $\alpha > 2$, we predict a sharp drop in the magic density, corresponding to a crossover from the ``Haar'' regime to the ``thermal'' regime at $|\beta| \approx \beta_c$, with $\beta_c$ vanishing exponentially in $N$. In the ``thermal'' regime, $\widetilde{M}_\alpha/N = (\alpha-1)^{-1} + \bigO{\beta^{2\alpha}}$ for small $\beta$ and $\alpha \geq 2$.}
    \label{fig:concept_fig}
\end{figure}

Our analysis yields a universal prediction for magic, quantified by the stabilizer R\'enyi entropy (SRE)~\cite{leone2022SRE,turkeshi2025fSRE}, with three main consequences. First, the SRE of equilibrium pure states is governed by a competition between two contributions to the Pauli spectrum: Haar-like fluctuations of high-weight Pauli strings, and thermal expectation values determined by $\sigma_\beta$. At infinite temperature the former dominate, recovering the Haar-value magic of mid-spectrum eigenstates~\cite{turkeshi2025fSRE}; at any finite temperature the latter take over, and energy conservation induces a volume-law thermodynamic correction to the SRE. Second, at high temperature the leading correction depends on the Hamiltonian only through the moments of its Pauli coefficients — a coarse thermodynamic ``fingerprint.'' Notably, the magic at Rényi index $\alpha$ is controlled by the $2\alpha$-th moment (for $\alpha \geq 2$), whereas classical thermodynamic quantities such as the free energy are controlled by the second: equilibrium magic at high temperature thus encodes information about the Hamiltonian invisible to ordinary thermodynamics. Third, for $\alpha > 2$ the competition between the two contributions becomes singular in the thermodynamic limit: the Haar regime survives only within a window $|\beta|\lesssim \beta_c$ that shrinks exponentially with system size, so that the filtered SRE (fSRE)~\cite{turkeshi2025fSRE} density is discontinuous at $\beta = 0$ [Fig.~\ref{fig:concept_fig}(b)]. In this sense, the infinite-temperature description of many-body magic is fragile, and the limits $N\to\infty$ and $\beta\to 0$ fail to commute.

Our work establishes magic as a thermodynamic property of chaotic quantum many-body systems. 
A direct physical implication of our work is that magic (as well as entanglement) in chaotic quantum many-body systems at sufficiently high temperature is long-range, i.e., cannot be removed by local finite-depth quantum circuits~\cite{ellison2021symmetry,korbany2025long,wei2025long,zhang2026extensive}.
While thermal Scrooge ensembles (and closely related ensembles~\cite{sugiura2013canonical}) have been proposed to describe other equilibrium properties~\cite{cotler2023projensemble,mark2024deeptherm,mok2026universalityscrooge,goldstein2006distribution,goldstein2016universal}, including entanglement~\cite{nakagawa2018universality}, our work suggests a unified framework for quantum resources in many-body physics.

\begin{figure*}[t!]
    \centering
\includegraphics[width=0.9\textwidth]{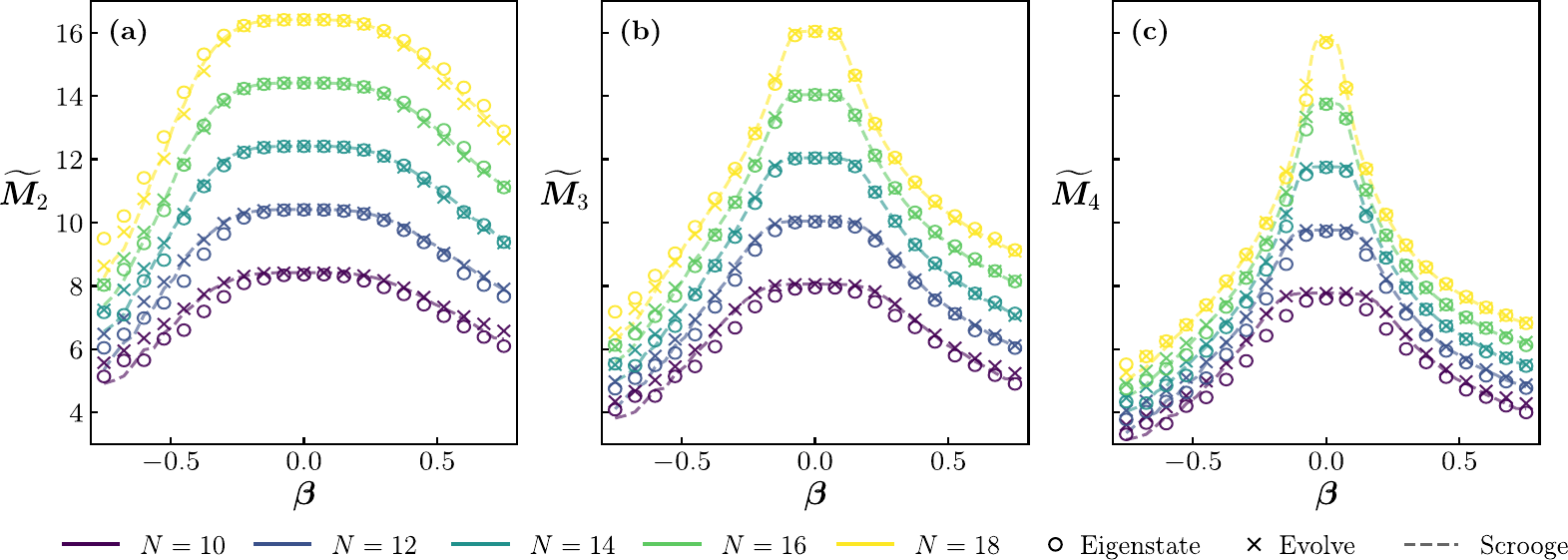}
    \caption{{Filtered stabilizer R\'{e}nyi entropy (fSRE) $\widetilde{M}_\alpha$ versus effective inverse temperature $\beta$ for (a) $\alpha = 2$, (b) $\alpha = 3$, and (c) $\alpha = 4$. We compare three classes of equilibrium pure states generated from the MFIM~\eqref{eq:fin_ham} with $10 \leq N \leq 18$: energy eigenstates, late-time evolved states obtained by evolving random product states to $Jt = 200$, and states sampled from Scrooge($\sigma_\beta$). As $N$ increases, the fSREs of the eigenstates and late-time evolved states agree closely with that of Scrooge($\sigma_\beta$).}}
    \label{fig:magic_beta}
\end{figure*}
\prlsection{Pauli spectrum, magic, and Scrooge ensembles} An arbitrary $N$-qubit quantum state $\rho$ in a $D = 2^N$ dimensional Hilbert space is fully characterized by its Pauli spectrum~\cite{beverland2020lower}
\begin{align}
    \ps(\rho) = \{\Tr(P \rho),~P \in \mathcal P_N\},
\end{align}
which is the set of all $D^2$ expectation values of Pauli strings $P = P_1 \otimes P_2 \otimes \cdots \otimes P_N$ where $P_i \in \{X,Y,Z,I\}$ and $\mathcal P_N$ is the Pauli group on $N$ qubits modulo phases. 

From the Pauli spectrum, one can define the stabilizer R\'enyi entropy (SRE)~\cite{leone2022SRE,haug2024SREalgs,oliviero2022measuring}
\begin{align}
M_\alpha(\rho) &= \frac{1}{1-\alpha}\log_2 \left(\frac{\ipr_\alpha(\rho)}{D}\right), \notag \\
\ipr_\alpha(\rho) &= \sum_{P \in \mathcal P_N} \left|\Tr(P\rho)\right|^{2\alpha}.\label{eq:sre_alpha}
\end{align}
Eq.~\eqref{eq:sre_alpha} holds for any $\alpha > 0$, with $M_1(\rho) \equiv \lim_{\alpha \to 1} M_\alpha(\rho)$.
The SREs $M_\alpha$ are magic monotones for pure states when $\alpha \geq 2$~\cite{leone2024stabilizer}, with $M_\alpha = 0$ for stabilizer states and $M_\alpha>0$ otherwise~\cite{Veitch2014ResourceTheoryStabilizer}.
$M_\alpha$ and $\ipr_\alpha$ can be thought of as the participation entropy and inverse participation ratio (IPR)~\cite{luitz2014ipr,sierant2022pe} in operator space, respectively~\cite{turkeshi2025fSRE,turkeshi2023opspace}. 
In this work, we will also use a slightly modified measure of magic, known as the filtered SRE (fSRE)~\cite{turkeshi2025fSRE}
\begin{align} \label{eq:fsre_alpha}
\fsre_\alpha (\rho) = \frac{1}{1-\alpha}\log_2 \left(\frac{\ipr_\alpha(\rho) - 1}{D-1}\right).
\end{align}
This definition removes the trivial identity contribution in $\zeta_{\alpha}$ and rescales the result so that $\fsre_\alpha = 0$ for stabilizer states. By eliminating this state-independent contribution, the fSRE is a more sensitive probe of magic~\cite{haug2026efficient}; in particular, it distinguishes typical highly scrambled states from atypical low-entanglement states~\cite{turkeshi2025fSRE}. This makes it well suited to chaotic quantum many-body systems, which are expected to obey quantum typicality~\cite{reimann2007typicality1,bartsch2009typicality2}. As we will show, the fSRE reveals crucial insights about the source of many-body magic which are absent in the SRE.

The central question of our work is: What is the magic of generic quantum many-body systems at finite energy densities relative to the ground state? Highly excited eigenstates of local chaotic many-body Hamiltonians near the middle of the spectrum have magic close to that of Haar random states~\cite{turkeshi2025fSRE}. Here, we propose a more general principle: many-body magic is universally captured by Scrooge ensembles. Scrooge ensembles generalize Haar random states by incorporating physical constraints. Among all ensembles consistent with those constraints, they are maximally informationally stingy. For a density matrix $\sigma$, the Scrooge ensemble is~\cite{jozsa1994scrooge,mark2024deeptherm,mok2026universalityscrooge}
\begin{align} \label{eq:cont_scr}
\text{Scrooge} (\sigma) = \left\{D\bra{\phi}\sigma\ket{\phi}\dd{\phi},\frac{\sqrt{\sigma}\ket \phi}{\norm{\sqrt{\sigma}\ket \phi}}\right\},
\end{align}
with $\dd \phi$ being the Haar measure on the unit sphere in $\mathbb C^{D}$. 

Next, we show that the Pauli spectrum of chaotic quantum many-body states is well described
by the thermal Scrooge ensemble, leading to a universal formula for many-body magic.

\prlsection{Universal behavior of magic} Consider a generic local chaotic many-body Hamiltonian $H$ with no conservation laws beyond energy, and equilibrium pure states $\ket{\psi}$ at finite energy density: eigenstates of $H$ or late-time states evolved from an initial product state. In both cases, quantum thermalization predicts that local observables are described by the thermal Gibbs state $\sigma_\beta$,
\begin{align} \label{eq:eth}
   \bra {\psi} O \ket {\psi} \approx \tr\parens{O\sigma_\beta},~\sigma_\beta = \frac{e^{-\beta H}}{\tr(e^{-\beta H})},
\end{align}
for any low-weight observable $O$~\cite{gogolin2016qtherm}. The effective inverse temperature $\beta$ is fixed by matching the energy to the canonical ensemble,
\begin{align} \label{eq:beta_energy}
  \braket{\psi|H|\psi} \equiv \tr\parens{H\sigma_\beta}.  
\end{align}
States with effective infinite temperature $(\beta = 0)$ lie near the middle of the spectrum of $H$; their magic was studied in Ref.~\cite{turkeshi2025fSRE}. We go beyond this limit by allowing $\beta = \bigO{1}$. The validity of Eq.~\eqref{eq:eth} is often formalized by the eigenstate thermalization hypothesis (ETH)~\cite{deutsch2018eigenstate,dalessio2016quantum}.

Magic, however, depends on the full Pauli spectrum $\Xi(\psi)$, including high-weight Pauli strings unconstrained by ETH. For a typical Scrooge state [Eq.~\eqref{eq:cont_scr}] with $\sigma = \sigma_\beta$ and a generic local Hamiltonian $H$, the distribution of $\braket{P}_\psi$ is illustrated in Fig.~\ref{fig:concept_fig}(a).
For finite $N$, the typical fSRE of Scrooge$(\sigma_\beta)$ for $\alpha > 2$ exhibits two regimes separated by a crossover inverse temperature $\beta_c$ (see End Matter for details), 
\begin{equation}
\label{eq:fSRE_scrooge}
    \widetilde{M}_\alpha(\psi) \approx
    \begin{cases}
        \displaystyle N - \frac{\log_2((2\alpha - 1)!!)}{\alpha - 1} + \bigO{2^{-N}},
        & |\beta| \lesssim \beta_c, \\[0.4em]
        \displaystyle \widetilde{M}_\alpha(\sigma_\beta) - \frac{\log_2(d_\alpha)}{\alpha-1},
        & |\beta| \gtrsim \beta_c.
    \end{cases}
\end{equation}
where $d_2 = 4$ and $d_\alpha = 1$ for integer $\alpha > 2$. For noninteger $\alpha$, we use the corresponding analytic continuation. The ``Haar'' regime is characterized by $|\beta| \lesssim \beta_c$, while $|\beta| \gtrsim \beta_c$ (with $\beta$ independent of $N$) gives the ``thermal'' regime [see Fig.~\ref{fig:concept_fig}(b)]. For $\alpha>2$, $\beta_c$ vanishes exponentially in $N$.
We propose Eq.~\eqref{eq:fSRE_scrooge} as the universal behavior of magic in generic quantum many-body systems, {supported by extensive numerical evidence (discussed below)}. Analogous results for the SRE ${M}_\alpha(\psi)$ are discussed in SM~\ref{sm:asymp_magic}.

As an illustration, we numerically study the mixed-field Ising model (MFIM), a paradigmatic chaotic many-body model~\cite{kim2013mfim}, though our theory is not specific to any Hamiltonian. We consider a 1D open spin-$1/2$ chain governed by
\begin{equation} \label{eq:fin_ham}
    H = \sum_{i=1}^N g_i X_i + \sum_{i=1}^N h_iZ_i + J\sum_{i=1}^{N-1}Z_iZ_{i+1} + H_1
\end{equation}
where $H_1 = Y_{N-2}Z_{N-1} + Y_{N-1}Z_N$ breaks time-reversal symmetry.
We use $h_1=J=1,~g_i = 1~\forall i$, and $h_i =0.5~\forall i \in \{2,\cdots,N\}$, for which $H$ has Gaussian Unitary Ensemble (GUE) spectral statistics~\cite{mehta2004random,haake2010quantum}. To reduce finite-size statistical fluctuations, we add random local magnetic fields to $H$ and average $\widetilde{M}_\alpha(\psi)$ over Hamiltonian instances (see End Matter).

Figure~\ref{fig:magic_beta} shows $\widetilde{M}_\alpha(\psi)$ for realistic examples of equilibrium pure states, such as energy eigenstates (computed using POLFED~\cite{sierant2020polynomially}) and late-time evolved states obtained by time-evolution of random product states, compared against the average magic of Scrooge$(\sigma_\beta)$. Here, $\beta$ is computed from Eq.~\eqref{eq:beta_energy}. The eigenstate and late-time magic agree excellently with Scrooge$(\sigma_\beta)$, even at modest system sizes, and show a plateau near $\beta = 0$ where $\widetilde{M}_\alpha \approx N$~\cite{turkeshi2025fSRE}; for $\alpha>2$, its width shrinks with $N$~\footnote{If $H$ obeys time-reversal symmetry, we observe analogous agreement with Eq.~\eqref{eq:cont_scr} using real-valued Haar random states.}. In the {SM}, we show further numerical evidence of agreement with Scrooge$(\sigma_\beta)$ for other local Hamiltonians, supporting the universality of our theory. Interestingly, our preliminary results suggest that magic may also serve as a probe of ergodicity, as the magic of quantum many-body scar eigenstates deviates sharply from the Scrooge value. A systematic study of scar-state magic is beyond the scope of this paper.

\begin{figure}
    \centering
    \includegraphics[width=0.9\columnwidth]{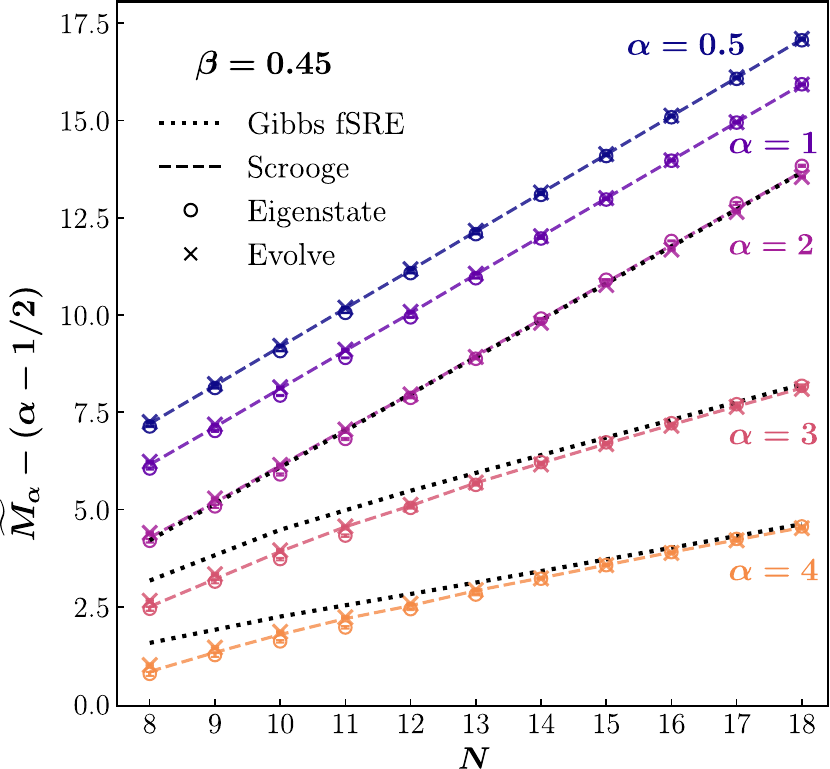}
    \caption{ {Filtered stabilizer R\'{e}nyi entropy (fSRE) $\widetilde{M}_\alpha$ versus number of qubits $N$ at $\beta = 0.45$, for $\alpha \in \{0.5, 1, 2, 3, 4\}$. A vertical offset of $\alpha - 1/2$ is added for visualization purposes. We compare energy eigenstates of the MFIM~\eqref{eq:fin_ham}, late-time evolved states obtained by evolving random product states to $Jt = 200$, and states sampled from Scrooge($\sigma_\beta$). The black dotted lines show $\widetilde{M}_\alpha(\sigma_\beta) - \log_2(d_\alpha)/(\alpha-1)$ for $\alpha \geq 2$ [Eq.~\eqref{eq:fSRE_scrooge}]. As $N$ increases, the fSREs of the various states agree closely with $\widetilde{M}_\alpha(\sigma_\beta)$.}}
    \label{fig:magic_N}
\end{figure}

Away from $\beta = 0$, Eq.~\eqref{eq:fSRE_scrooge} predicts $\widetilde{M}_\alpha(\psi) \approx \widetilde{M}_\alpha(\sigma_\beta)$ up to a constant. Figure~\ref{fig:magic_N} verifies this for the MFIM and $\alpha \geq 2$, with convergence at sufficiently large $N$. Better convergence at smaller $N$ is observed for $\alpha = 2$, due to the lack of a sharp crossover between the ``Haar'' and ``thermal'' regimes [cf. Fig.~\ref{fig:concept_fig}(b)].

\begin{figure*}[t!]
    \centering
    \includegraphics[width=0.9\textwidth]{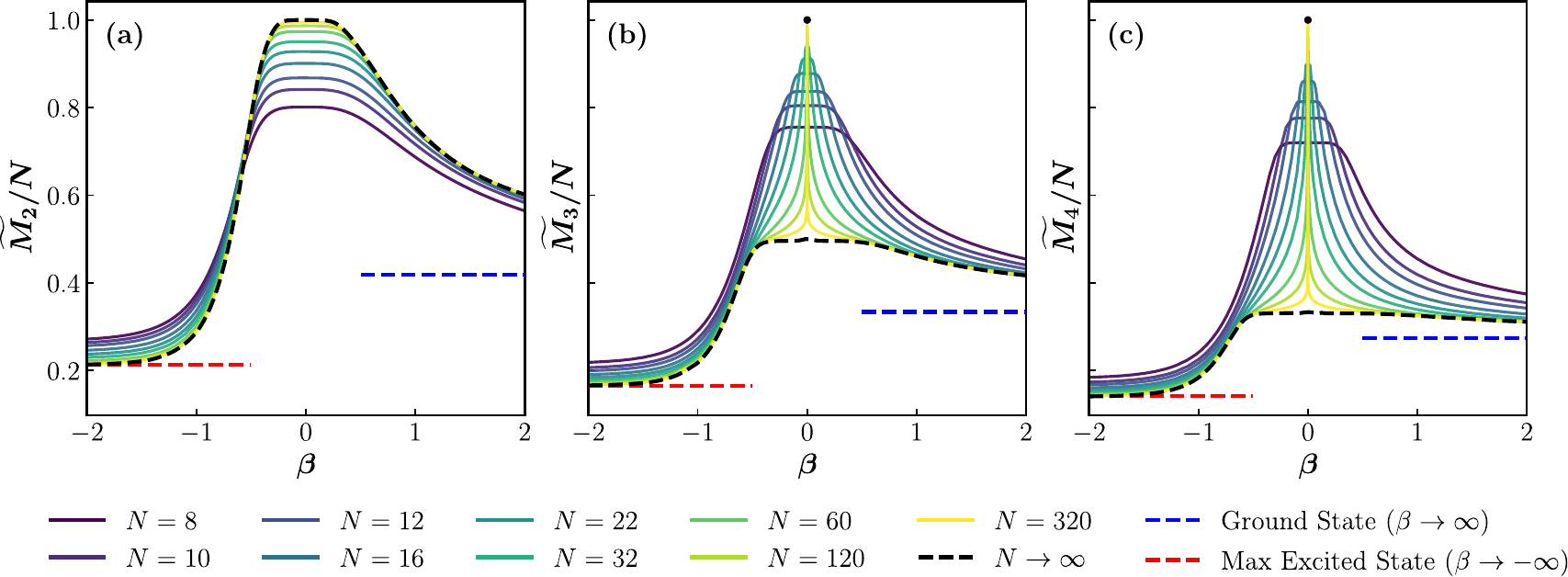}
    \caption{{Filtered stabilizer R\'{e}nyi entropy (fSRE) density $\widetilde{M}_\alpha/N$ versus effective inverse temperature $\beta$ for (a) $\alpha = 2$, (b) $\alpha = 3$, and (c) $\alpha = 4$. The fSRE is calculated using the approximation $\ipr_\alpha(\psi) \approx 2^{N(2-\alpha)} (2\alpha - 1)!! + d_\alpha \ipr_\alpha(\sigma_\beta)$ inspired by Eq.~\eqref{eq:scrooge_ipr_main}, capturing both ``Haar'' and ``thermal'' regimes. The black dashed line shows the $N \to \infty$ limit of the fSRE (and also the SRE) density, estimated by extrapolating numerical results for $8 \leq N \leq 320$; blue and red lines mark the limiting fSRE densities of the ground state and maximally excited state, respectively. For $\alpha > 2$, the black circle marks the infinite-temperature value $\widetilde{M}_\alpha/N = 1$.}}\label{fig:magic_density}
\end{figure*}

That the fSRE is captured so well at high temperature by Scrooge$(\sigma_\beta)$ is rather surprising. Scrooge$(\sigma_\beta)$ represents canonical typicality, which is appropriate for describing local properties such as local observables and projected ensembles~\cite{canonical_typicality_2024,cotler2023projensemble}. In contrast, our equilibrium states -- whether Hamiltonian eigenstates or late-time-evolved states -- are associated with microcanonical rather than canonical ensembles. These states ought to agree with Scrooge$(\sigma_\beta)$ on low-weight observables but need not agree on high-weight observables. Magic, however, is a global feature that depends on all Pauli strings. Why then does the canonical Scrooge ensemble capture this global property?

The resolution of this puzzle is that the IPR $\ipr_\alpha$ (see Eq.~\eqref{eq:sre_alpha}), though a sum over all Pauli strings, is dominated at high temperature by local structure. Consider an $N$-qubit \emph{local} Hamiltonian $H = \sum_{Q \in \mathcal{P}_N} a_Q Q$ with $\bigO{N}$ nonzero terms, each with weight $\bigO{1}$; suppose $H$ is normalized such that $\sum_{Q} |a_Q|/N = 1$, and that the energy spectrum is shifted such that $\Tr(H)=0$ and hence $a_I=0$. A high-temperature cluster expansion~\cite{brydges1986short,oitmaa2006series} (see End Matter for details) yields
\begin{equation} \label{eq:cluster_exp}
    \ln \ipr_\alpha(\sigma_\beta) = \beta^{2\alpha} \sum_{Q \in \mathcal{P}_N} {|a_Q|^{2\alpha}} + \bigO{N\beta^{2\alpha + 1}},
\end{equation}
where $\alpha \geq 2$ is a positive integer. The leading term for small $\beta$ comes from the low-weight Pauli operators in $H$ itself, while higher-order terms arise from products of these operators, namely connected clusters with diameter comparable to the thermal correlation length. For equilibrium pure states, these clusters are low-weight operators and therefore governed by subsystem ETH~\cite{dymarsky2018subsystem,garrison2018does}, explaining why $\ln \ipr_\alpha$ inherits its value from $\sigma_\beta$ even though high-weight Pauli strings may not. Two observations then follow. First, near a thermal critical point, where the correlation length diverges, our theory may break down. Second, locality is crucial: for a nonlocal Hamiltonian such as a matrix drawn from GUE, eigenvectors are Haar random and $\widetilde{M}_\alpha(\psi)$ becomes $\beta$-independent rather than following $\widetilde{M}_\alpha(\sigma_\beta)$.

\prlsection{Magic density in the thermodynamic limit} 
Combining Eq.~\eqref{eq:fSRE_scrooge} with the high-temperature cluster expansion [Eq.~\eqref{eq:cluster_exp}], we predict that, for $\alpha \geq 2$ and $N \to \infty$,
\begin{equation}
\label{eq:fsre_density}
    \frac{\widetilde{M}_\alpha(\psi)}{N} =
    \begin{cases}
        \displaystyle 1,
        & \beta = 0, \\[0.6em]
        \displaystyle \frac{1}{\alpha - 1}
        \sparens{1 - c_\alpha \beta^{2\alpha} + \bigO{\beta^{2\alpha + 1}}},
        & \beta \neq 0.
    \end{cases}
\end{equation}
For $\alpha < 2$ and $N \to \infty$, we perform a Taylor series expansion around $\beta = 0$ as an approximation of the Scrooge magic to get 
\begin{align}
  \frac{\fsre_\alpha(\psi)}{N} = 1-\frac{\alpha c_2}{2}\beta^4 + \bigO{\beta^5},  
\end{align}
where $c_\alpha \equiv (N \ln 2)^{-1} \sum_{Q} {|a_Q|}^{2\alpha}$ is $\bigO{1}$ for local Hamiltonians. Details of the derivation are presented in SM~\ref{sm:asymp_magic}. For $\beta \neq 0$, energy conservation induces a volume-law thermodynamic correction to magic. At high temperature, the leading correction depends on $H$ only through $c_\alpha$, which quantifies the delocalization of $H$ in the Pauli basis. Thus the Hamiltonian leaves a coarse thermodynamic fingerprint on equilibrium magic, enabling comparisons through the $\alpha$-dependence of magic. This also distinguishes magic from ordinary thermodynamics: $c_1$ controls the high-temperature correction to the free energy, $\ln Z_\beta = N \ln 2 \parens{1 + c_1 \beta^2 / 2 + \bigO{\beta^3}}$, whereas the fSRE at R\'{e}nyi index $\alpha \geq 2$ is controlled by $c_\alpha$. This is consistent with the fact that magic is a genuinely quantum resource with no classical analogue.

For $\alpha > 2$, Eq.~\eqref{eq:fsre_density} gives a discontinuous transition in magic density at $\beta = 0$, from $1$ to $1/(\alpha - 1)$ [Fig.~\ref{fig:concept_fig}(b)]. Thus the infinite-temperature theory of many-body magic~\cite{turkeshi2025fSRE} is fragile against finite-temperature corrections, while the more general framework using Scrooge$(\sigma_\beta)$ captures the correct behavior at both finite and infinite temperatures. This transition is not visible in the SRE, reinforcing the point that the fSRE is a more sensitive diagnostic of magic. Physically, $\ipr_\alpha$ in Eq.~\eqref{eq:scrooge_ipr_main} is dominated by Gaussian-like fluctuations for $|\beta| \lesssim \beta_c$, and by thermal averages for $|\beta| \gtrsim \beta_c$; since $\beta_c$ is exponentially small in $N$, this becomes a singularity in the thermodynamic limit. Direct verification for MFIM eigenstates and late-time evolved states is inaccessible at such large $N$, so Fig.~\ref{fig:magic_density} shows the expected behavior up to $N=320$, calculated using the approximation $\ipr_\alpha(\psi) \approx 2^{N(2-\alpha)} (2\alpha - 1)!! + d_\alpha \ipr_\alpha(\sigma_\beta)$ with $\ipr_\alpha(\sigma_\beta)$ computed using a tensor network algorithm~\cite{haug2023nonstab,tarabunga2024nonstabilizerness} (see End Matter).

\prlsection{Discussion and outlook} We have proposed a theory of magic in generic quantum many-body systems. For non-integrable systems, equilibrium pure states such as highly excited eigenstates and late-time evolved states are described by the thermal Scrooge ensemble, in excellent agreement with extensive numerical simulations of the MFIM. For systems with no conservation laws beyond energy, equilibrium magic is universally described by $\widetilde{M}_\alpha(\sigma_\beta)$, depending only on the effective inverse temperature $\beta$. Energy conservation therefore induces a volume-law thermodynamic correction to magic. This parallels the relation between entanglement and thermodynamic entropy and complements work showing that entanglement entropy is captured by the same thermal Scrooge ensemble~\cite{nakagawa2018universality} (additional numerics shown in SM~\ref{sm:supp_numerics}), suggesting a unified view of many-body quantum resources. 

Several physical implications follow. First, although SRE and fSRE are not mixed-state magic monotones~\cite{haug2026efficient,tarabunga2025efficient,ding2025evaluating}, Eq.~\eqref{eq:fSRE_scrooge} endows $\widetilde{M}_\alpha(\sigma_\beta)$ (and, similarly, $M_\alpha(\sigma_\beta)$) with a concrete physical meaning: it describes the magic of pure many-body states at thermal equilibrium, not the magic of the Gibbs state itself.

Second, high-temperature many-body states have long-range magic~\cite{ellison2021symmetry,korbany2025long,wei2025long,zhang2026extensive} robust against geometrically local constant-depth circuits.
To reach this second conclusion, in SM~\ref{sm:robustness} we compare $c_\alpha(H)$ with $c_\alpha(UHU^\dagger)$, where $H$ is a $k$-local Hamiltonian and $U$ is a $d$-dimensional geometrically local depth-$t$ unitary circuit. We find that conjugating $H$ by $U$ increases $c_\alpha$ by at most a factor $\bigO{t^{d(\alpha - 1)}}$; therefore, Eq.~\eqref{eq:fsre_density} shows that $U$ induces a small relative change in $\widetilde{M}_\alpha$ for $\beta \ll t^{-d(\alpha - 1)/(2\alpha)}$. This immediately implies that high-temperature equilibrium pure states are long-range entangled~\cite{korbany2025long}.

Finally, combined with Ref.~\cite{mok2026universalityscrooge}, our results imply that high-temperature many-body systems contain the magic required for deep thermalization when the bath is measured in a typical Clifford basis~\cite{loio2026quantum, bittel2026operational}.

We anticipate that this work will provide a foundation for a systematic understanding of the connection between magic and quantum many-body systems. We expect the theory to be broadly applicable beyond the scope of the present work, leaving several directions for future investigation. For example, if other local conserved quantities are present, such as charge in systems with U(1) symmetry~\cite{iannotti2026nonstabilizerness}, it is likely that magic is also well described by Scrooge ensembles, with $\sigma_\beta$ replaced by a Gibbs state depending on both temperature and other thermodynamic parameters such as chemical potential. Extending the theory to non-chaotic systems would be another interesting direction. For instance, the magic of integrable many-body systems may be described by the Scrooge ensemble associated with the generalized Gibbs ensemble~\cite{rigol2007gge1,vidmar2016gge2}, thereby establishing a common framework across a broad class of complex many-body systems.

\begin{acknowledgments}
\prlsection{Acknowledgments}
We thank Piotr Sierant for providing the code for POLFED~\cite{sierant2020polynomially,pintar2026computing} and fast SRE computation~\cite{sierant2026computing}. Portions of the code used for simulation were developed with assistance from OpenAI's Codex, and thoroughly reviewed by the authors. J.~P.~acknowledges support from the U.S. Department of Energy, Office of Science, National Quantum Information Science Research Centers, Quantum Systems Accelerator, and the National Science Foundation (PHY-2317110). The Institute for Quantum Information and Matter is an NSF Physics Frontiers Center.
\end{acknowledgments}

\bibliography{bib}

\newpage
\vspace{2cm}
\begin{center}
\textbf{\large End Matter}
\end{center}

\section{Analytical expressions for Scrooge-averaged magic}

To obtain an analytical approximation of the average fSRE over Scrooge$(\sigma_\beta)$, we analytically compute the annealed fSRE average~\cite{sierant2026theorymagicphasetransitions},
\begin{equation}
    \widetilde{M}_{\alpha,\text{annealed}} = \frac{1}{1-\alpha} \log_2 \left(\frac{\E_{\psi}\ipr_\alpha(\psi) - 1}{D-1}\right),
\end{equation}
where $\E_{\psi} \equiv \E_{\psi \sim \text{Scr.}(\sigma_\beta)}$. $\text{Scr.}(\sigma_\beta)$ is the thermal Scrooge ensemble. This is more analytically tractable to compute than the actual Scrooge-averaged fSRE (also known as the quenched average). We have numerically checked the validity of the annealed fSRE as an approximation of the average fSRE. This approximation is only used for analytical arguments; the numerical results presented in the main text are for the actual Scrooge-averaged fSRE.
Computing this Scrooge average for $\alpha \geq 2$ gives (see SM~\ref{sm:scrooge_magic} for details)
\begin{equation} \label{eq:scrooge_ipr_main}
    \E_{\psi}[\ipr_\alpha(\psi)] \approx
    \begin{cases}
        \displaystyle 2^{N(2-\alpha)} (2\alpha - 1)!!+1,
        & N \text{ fixed},\ \beta \to 0, \\[0.6em]
        \displaystyle d_\alpha\ipr_\alpha(\sigma_\beta),
        & N \to \infty,\ \beta \text{ fixed}
    \end{cases}
\end{equation}
where $d_2 = 4$, while $d_\alpha = 1$ for integer $\alpha > 2$. Our derivation uses an unnormalized deformed-Haar approximation to the Scrooge ensemble~\cite{mark2024deeptherm} for analytical tractability, which is a good approximation to Scrooge($\sigma_\beta$) for low-purity $\sigma_\beta$~\cite{mcginley2025scrooge,mok2026universalityscrooge}. The averaged IPR is expressed in terms of a sum over permutation-cycle contributions involving $t_m(P) = \Tr\sparens{(P\sigma_\beta)^m}$. The identity permutation gives $(t_1(P))^{2\alpha}$, whose sum over $P$ yields $\ipr_\alpha(\sigma_\beta)$. For integer $\alpha > 2$, the contributions from other permutations are bounded using the generalized purities of $\sigma_\beta$, which are typically exponentially small in $N$ at sufficiently high fixed temperature. This gives $d_\alpha = 1$, as only the identity permutation contributes in the thermodynamic limit. For $\alpha = 2$, other permutations also contribute to leading order, giving $d_2 = 4$. This yields our approximation~\eqref{eq:fSRE_scrooge} for the fSRE. The first case is the previously studied ``Haar'' regime~\cite{turkeshi2025fSRE}, where $\ipr_\alpha(\psi)$ is dominated by Haar-random fluctuations in $\braket{P}_\psi$, with $|\braket{P}_\psi| \sim 2^{-N/2}$~\cite{turkeshi2025fSRE}; the factor $(2\alpha - 1)!!$ reflects their Gaussianity~\footnote{This remains valid for non-integer $\alpha > 2$, with the double factorial replaced by $2^\alpha \Gamma(\alpha + 1/2)/\sqrt{\pi}$.}. The second is the finite-temperature ``thermal'' regime, dominated by $\E \braket{P}_\psi = \Tr(P \sigma_\beta)$. Given that we write the Hamiltonian as $H = \sum_{Q \in \mathcal P_N} a_Q Q$, matching the two asymptotics at finite $N$ gives
\begin{align} \label{eq:betac}
    \beta_c \approx \left(\frac{(2\alpha - 1)!!}{d_\alpha c_\alpha\ln2}\right)^{\frac{1}{2\alpha}} N^{-\frac{1}{2\alpha}}2^{-N{\frac{\alpha-2}{2\alpha}}},
\end{align}
where $c_\alpha = (N\ln2)^{-1}(\sum_Q |a_Q|^{2\alpha})$ is an $\bigO{1}$ constant. We have also numerically checked that the fSRE values of Scrooge$(\sigma_\beta)$ states concentrate about their ensemble average, with relative fluctuations vanishing exponentially with system size. Self-averaging provides the basis for interpreting Eq.~\eqref{eq:fSRE_scrooge} as the fSRE of a typical state drawn from Scrooge$(\sigma_\beta)$, since the fluctuations become negligible in the thermodynamic limit. Although we do not prove self-averaging for the fSRE, we demonstrate this in SM~\ref{sm:ipr_cond} for a related measure of magic known as the linearized fSRE~\cite{haug2024SREalgs}.

\section{High-temperature cluster expansion of Gibbs IPR}

{As stated in the main text, we write a general $N$-qubit Hamiltonian as $H = \sum_{P \in \mathcal{P}_N} a_P P$, with $a_I = 0$ without loss of generality. For a fixed integer $\alpha \geq 2$, we compute the high-temperature expansion of $\ln \ipr_\alpha(\sigma_\beta)$ for the Gibbs state $\sigma_\beta = e^{-\beta H}/Z_\beta$:
\begin{equation}
    \ln \ipr_\alpha(\sigma_\beta) = \ln \Tr\parens{Q_\alpha e^{-\beta H_R}} - 2\alpha \ln Z_\beta,
\end{equation}
where
\begin{equation}
    Q_\alpha = \sum_{P \in \mathcal{P}_N} P^{\otimes 2\alpha} = (I^{\otimes 2\alpha} + X^{\otimes 2\alpha} + Y^{\otimes 2\alpha} + Z^{\otimes 2\alpha})^{\otimes N},
\end{equation}
and
\begin{equation}
    H_R \equiv \sum_{r=1}^{2\alpha} H^{(r)} = \sum_{r=1}^{2\alpha} I^{\otimes (r-1)} \otimes H \otimes I^{\otimes (2\alpha-r)}.
\end{equation}
The operator $H_R$ acts on $2\alpha$ replicas of the system. Defining the normalized expectation values $\braket{\boldsymbol{\cdot}}_0 \equiv \Tr(\boldsymbol{\cdot})/2^N$ and $\braket{\boldsymbol{\cdot}}_Q \equiv \Tr(\boldsymbol{\cdot} Q_\alpha)/\Tr(Q_\alpha) = \Tr(\boldsymbol{\cdot} Q_\alpha)/2^{2\alpha N}$, we obtain the cumulant expansion
\begin{align}
\label{eq:cumulant_expansion}
    \ln \ipr_\alpha(\sigma_\beta)
    &= \sum_{n=1}^{\infty} \frac{(-\beta)^n}{n!}
    \biggl[\kappa_n^{(Q)}(H_R,\ldots,H_R) \nonumber \\
    &\hspace{4.5em}{}-2\alpha\kappa_n^{(0)}(H,\ldots,H)\biggr],
\end{align}
where $\kappa_n^{(0)}$ and $\kappa_n^{(Q)}$ are the $n$-th order cumulants with respect to $\braket{\boldsymbol{\cdot}}_0$ and $\braket{\boldsymbol{\cdot}}_Q$, respectively. Substituting the Pauli expansion of $H$, we can write
\begin{align}
    \ln \ipr_\alpha(\sigma_\beta)
    &= \sum_{n=1}^{\infty} \frac{(-\beta)^n}{n!}
    \sum_{P_1,\ldots,P_n} a_{P_1}\cdots a_{P_n} \nonumber \\
    &\quad\times\biggl[\kappa_n^{(Q)}(P_{1,R},\ldots,P_{n,R}) \nonumber \\
    &\hspace{5.5em}{}-2\alpha\kappa_n^{(0)}(P_1,\ldots,P_n)\biggr],
\end{align}
where $P_R = \sum_{r=1}^{2\alpha} I^{\otimes (r-1)} \otimes P \otimes I^{\otimes (2\alpha-r)}$. The cumulants $\kappa_n^{(0)}$ and $\kappa_n^{(Q)}$ can be nonzero only if the Pauli operators $P_1,\ldots,P_n$ form a connected cluster. The first nonvanishing contribution to $\ln \ipr_\alpha(\sigma_\beta)$ occurs at $n=2\alpha$. Indeed, for every nonidentity Pauli operator $P$, $\Tr(Pe^{-\beta H})=\bigO{\beta}$ as $\beta\to0$. It follows that $\ipr_\alpha(\sigma_\beta)=1+\bigO{\beta^{2\alpha}}$ and hence $\ln \ipr_\alpha(\sigma_\beta)=\bigO{\beta^{2\alpha}}$. The direct expansion gives (see SM~\ref{sm:gibbs_ipr})
\begin{equation}
    \ln \ipr_\alpha(\sigma_\beta) = \beta^{2\alpha} \sum_{P \neq I} |a_P|^{2\alpha} + \bigO{N\beta^{2\alpha + 1}}.
\end{equation}
For a finite-range, bounded-strength local Hamiltonian, the number of nonzero $a_P$ is $\bigO{N}$, so the leading coefficient is extensive. If, in addition, the linked-cluster expansion of $N^{-1}\ln \ipr_\alpha(\sigma_\beta)$ converges for $|\beta|<\beta_0$, with $\beta_0>0$ independent of $N$, then the fSRE density (i.e., the magic density) is well defined in the thermodynamic limit. The cluster expansion~\eqref{eq:cumulant_expansion} implies that higher-order thermal contributions to magic arise from connected clusters of operators of increasing size, enforcing an effective locality. 
}

\section{Numerical methods}

{To compute the effective inverse temperature $\beta$ for a given equilibrium pure state, we first compute the map $E(\beta) = \Tr(H \sigma_\beta)$ for various values of $\beta$. This is computed using exact diagonalization (ED) for $N \leq 14$ to obtain all the eigenenergies of $H$. For $14 \leq N \leq 18$, we use a kernel polynomial method via a truncated Chebyshev expansion~\cite{weise2006kpm} to calculate $E(\beta)$. Inverting this map then gives $\beta$ for a given energy $E = \braket{H}_\psi$ corresponding to the pure state $\ket{\psi}$. For $14 \leq N \leq 18$, we use POLFED~\cite{sierant2020polynomially} to sample energy eigenstates near a target $\beta$. The SRE and fSRE are calculated using the HadaMAG algorithm~\cite{sierant2026computing}, which is based on the fast Walsh--Hadamard transform.}

To mitigate statistical fluctuations inherent to small system sizes, we disorder-average over an ensemble of Hamiltonians. This is achieved by randomizing the local longitudinal and transverse fields of the MFIM in Eq.~\eqref{eq:fin_ham}, drawing the field strengths $h$ and $g$ uniformly from the intervals $[0.35, 0.65]$ and $[0.85, 1.15]$, respectively. For a given target inverse temperature $\beta_0$, the magic value for each randomly sampled Hamiltonian is computed at an inverse temperature closest to $\beta_0$. For $N=18$, we checked that the finite-size fluctuations are sufficiently small such that this ensemble averaging is no longer necessary, and a single sample suffices.

To sample from Scrooge($\sigma_\beta$), the key step is to apply $\sqrt{\sigma_\beta} = \exp(-\beta H/2)$ to a Haar-random state $\ket{\phi}$. Normalizing and adjusting the weights then yields samples of the Scrooge ensemble. For $N \geq 14$, constructing the full $2^N \times 2^N$ matrix for $\sqrt{\sigma_\beta}$ is impractical. Therefore, we expand $\exp(-\beta H/2)$ as a linear combination of Chebyshev polynomials $T_m(H)$, truncated at order $M$. Since $H$ is a sparse matrix, the action of $T_m(H)$ on $\ket{\phi}$ can be computed using sparse matrix-vector multiplication~\cite{weise2006kpm}. Numerical accuracy is validated by increasing the truncation $M$ and checking that the results have converged.

We remark that analogous results are obtained for Hamiltonians exhibiting time-reversal symmetry, with Scrooge($\sigma_\beta$) constructed by distorting real-valued Haar-random states instead of complex-valued Haar-random states.

\section{Tensor network algorithm for computing Gibbs IPR}

To compute $\ipr_\alpha(\sigma_\beta)$ numerically for $N$ up to $320$ in Fig.~\ref{fig:magic_density}, we use the tensor network algorithm adapted from Ref.~\cite{haug2023nonstab}, which computes the magic of pure states represented by matrix product states (MPS) (see also Ref.~\cite{tarabunga2024nonstabilizerness}). The main idea is to observe that, for positive integer $\alpha$,
\begin{equation}
    \ipr_\alpha(\sigma_\beta) = \sum_{P \in \mathcal{P}_N} \sparens{\Tr(P \sigma_\beta)}^{2\alpha} =  \Tr\parens{Q_\alpha \sigma_\beta^{\otimes 2\alpha}},
\end{equation}
where
\begin{equation}
    Q_\alpha = \sum_P P^{\otimes 2\alpha} = (I^{\otimes 2\alpha} + X^{\otimes 2\alpha} + Y^{\otimes 2\alpha} + Z^{\otimes 2\alpha})^{\otimes N}
\end{equation}
factorizes into local qubit operators acting on $2\alpha$ replica copies. $e^{-\beta H}$ is constructed as a matrix product operator (MPO) using the time-evolving block decimation (TEBD) algorithm in imaginary time. Then, $\ipr_\alpha(\sigma_\beta)$ can be computed by contracting $Q_\alpha$ locally with the MPO and normalizing by the partition function $Z_\beta$. If $\chi$ is the maximum bond dimension of the vectorized MPO for $e^{-\beta H}$, the resulting replica MPS has bond dimension $\chi^{2\alpha}$, and the local transfer matrix has size $\chi \times \chi$. Thus, the tensor network contraction step requires runtime $\bigO{N \alpha \chi^{2\alpha + 1}}$ and memory $\bigO{\chi^{2\alpha}}$. The code is implemented in Python with assistance from OpenAI's Codex and was thoroughly reviewed by the authors. In our numerical simulations, we used a TEBD step size of $0.05$, an SVD truncation cutoff of $10^{-10}$, $\chi = 12$ for $\alpha = 2, 3$, and $\chi = 8$ for $\alpha = 4$, which we verified to be sufficient for convergence of the results.

\clearpage

\let\addcontentsline\oldaddcontentsline

\makeatletter
\let\oldaddtocontents\addtocontents
\renewcommand{\addtocontents}[2]{}
\appendix
\let\addtocontents\oldaddtocontents
\makeatother
\onecolumngrid

\begin{center}

\textbf{\large Supplemental Material}
\end{center}

\setcounter{secnumdepth}{2}
\renewcommand{\thesection}{\Alph{section}}
\renewcommand{\thesubsection}{\arabic{subsection}}
\renewcommand*{\theHsection}{\thesection}
\renewcommand\appendixname{\SM{}}

In the Supplemental Material, we provide additional proofs and results to support the main text.
\makeatletter
\begingroup
\let\appendix\relax
\@starttoc{toc}
\endgroup
\makeatother

\section{Pauli spectra across different $\beta$}

\begin{figure}
    \centering
    \includegraphics[width=\linewidth]{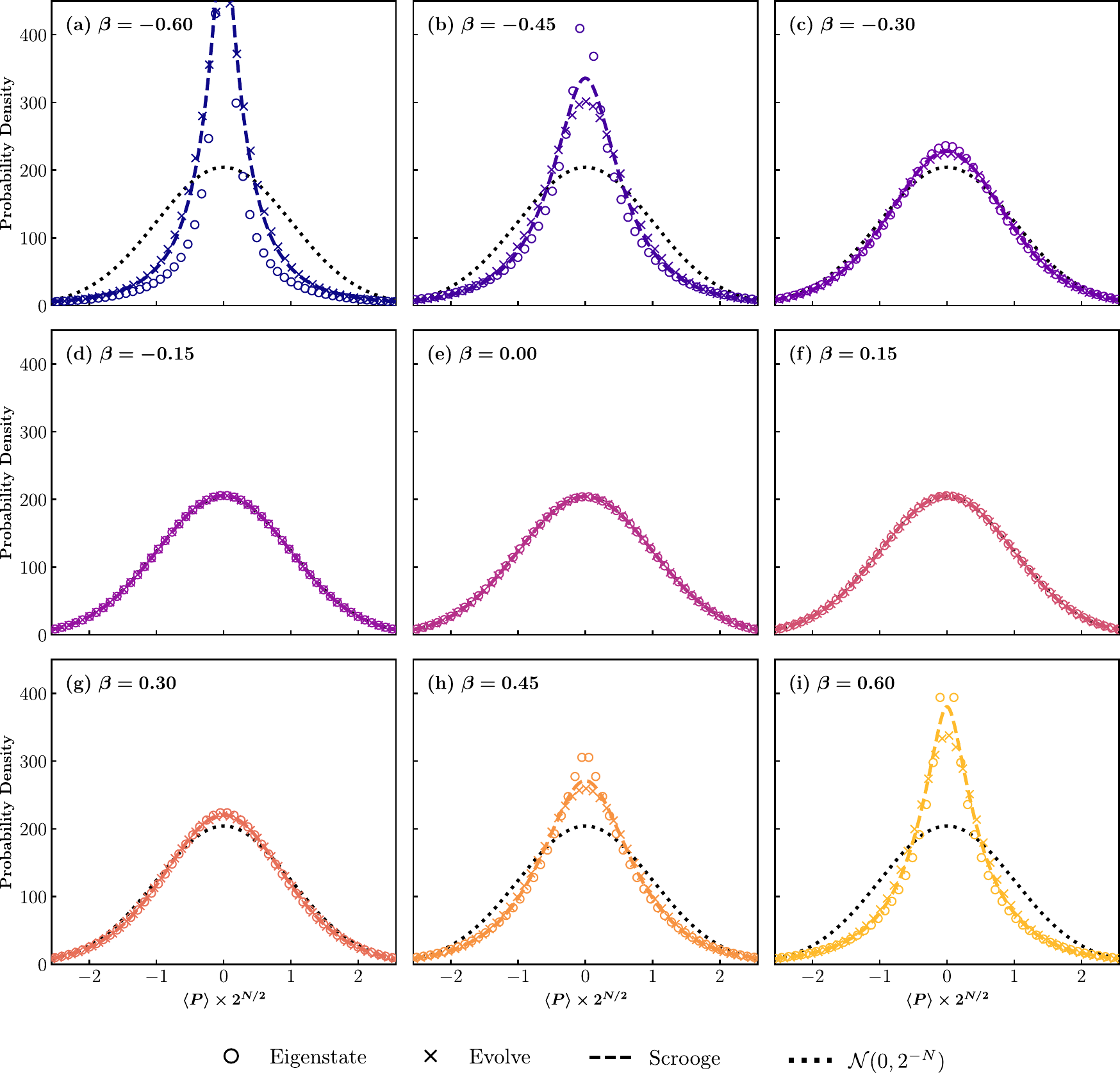}
    \caption{Pauli spectrum distributions for eigenstates of the MFIM Hamiltonian, thermal Scrooge states, and late-time-evolved states for various inverse temperatures $\beta \in [-0.6, 0.6]$. The distribution deviates from the typical Gaussian distribution (denoted by the black dashed line) corresponding to the Pauli spectrum of a Haar-random state (i.e., $\beta = 0$) as we move away from infinite temperature.}
    \label{fig:pauli_specs}
\end{figure}

In Fig.~\ref{fig:pauli_specs}, we present the probability density function of the rescaled Pauli expectation values, $x = \langle P \rangle \times 2^{N/2}$, across a range of inverse temperatures $\beta$. This rescaling ensures that the second moment of the distribution over all $4^N$ Pauli operators remains exactly unity. At infinite temperature ($\beta = 0$), the Pauli spectrum for all numerical methods closely follows a standard Gaussian distribution. This behavior is characteristic of typical Haar-random states~\cite{turkeshi2025fSRE}. However, as the system moves away from infinite temperature (increasing $\vert{}\beta\vert{}$), the distributions become non-Gaussian and sharper around 0.

\begin{figure}
    \centering
    \includegraphics[width=0.5\linewidth]{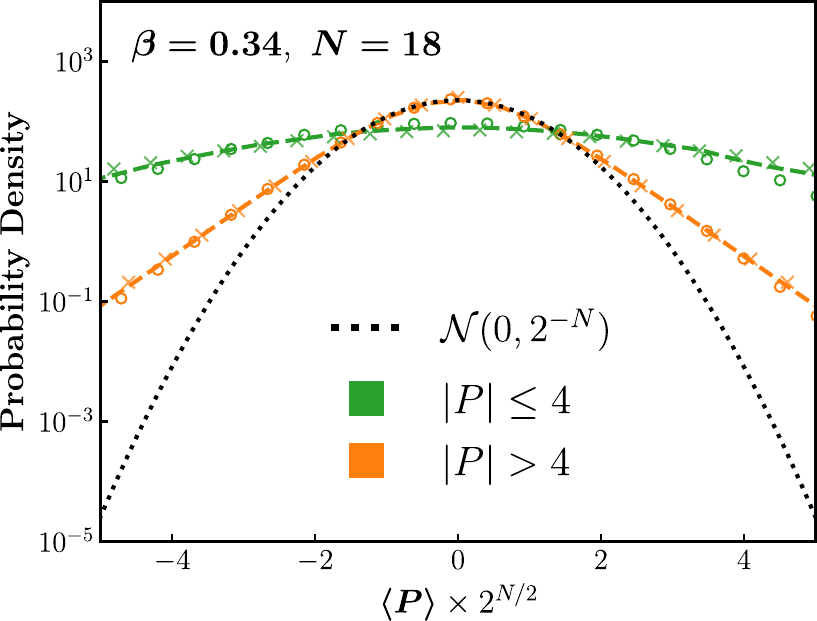}
    \caption{Pauli spectrum distributions for low-weight ($|P| \le |P|_c$) and high-weight ($|P| > |P|_c$) Paulis, with $|P|_c = 4$, chosen arbitrarily for visualization purposes. We see that low-weight Paulis behave thermally, while high-weight Paulis typically obey a Gaussian distribution (denoted by the dotted black curve). Possible deviations from the Gaussian distribution for high-weight Pauli operators may arise from the inclusion of thermally behaving high-weight Paulis formed by disjoint products of ``thermal-like'' low-weight Pauli operators.}
    \label{fig:lowhigh_specs}
\end{figure}

This can be explained using the observations in Fig.~\ref{fig:lowhigh_specs}, which highlight the contributions to the Pauli spectrum from high- and low-weight Pauli operators at a finite temperature. Since $N = 18$, we arbitrarily choose a cutoff weight $|P|_c = 4$. We note that the low-weight Pauli operators acquire thermal expectation values from subsystem ETH, while most high-weight Pauli operators follow a Gaussian distribution. We expect some high-weight Pauli operators, which are disjoint products of low-weight Pauli operators, to behave like products of thermal expectation values. However, these constitute an exponentially small fraction of the high-weight Pauli operators and do not affect the Gaussian profile. The appropriately normalized weighted sum of these two profiles gives the final non-Gaussian distribution, with a sharp peak around zero and longer tails.

\section{Self-averaging of the linearized SRE}
\label{sm:ipr_cond}

In this section, we show the conditions for concentration of measure (i.e., self-averaging) in the linearized SRE for Scrooge states $M^{\rm lin}_\alpha(\psi) = 1 - \xi_\alpha(\psi)$, where $\xi_\alpha(\psi) = \ipr_\alpha(\psi)/D = \sum_{P} \braket{\psi|P|\psi}^{2\alpha}/D$; namely,
\begin{align}
 \Eset{\psi \sim \text{Scr.}(\sigma)} [M^{\rm lin}_\alpha(\psi)]  \approx  M^{\rm lin}_\alpha(\psi),
\end{align}
with the approximation becoming an equality in the thermodynamic limit. For this, we have to prove self-averaging for the normalized IPR $\xi_\alpha$. We use the following theorem from Ref.~\cite{canonical_typicality_2024}, which is L\'evy's lemma for GAP measures,
\begin{align}\label{eq:levy_lemma}
  \mathbb P\left[\left|f(\psi) - \Eset{\psi \sim \text{Scr.}(\sigma)}(f)\right| \ge \epsilon\right] \le 6\exp{\left(\frac{-C\epsilon^2}{\eta^2 \norm{\sigma}}\right)},
\end{align}
where $f$ is a Lipschitz continuous function with $\eta$ as the Lipschitz constant. $C = 1/288\pi^2$ and $\norm{\sigma}$ is the operator norm of $\sigma$. Since $\xi_{\alpha}$ is a continuous function, we can directly use Eq.~\eqref{eq:levy_lemma} as long as we are able to calculate the Lipschitz constant, which is defined as
\begin{align}
 \eta = \max_{\|\psi\|=1} \| \nabla f(\psi) \|.
\end{align}
We can take the gradient statevector as
\begin{align}
 \ket{\nabla \xi_{\alpha}(\psi)} &= \frac{2\alpha}{D}\sum_P\braket{\psi|P|\psi}^{2\alpha-1} P\ket{\psi} \notag \\
 \implies \norm{\ket{\nabla \xi_{\alpha}(\psi)}} &\le \frac{4\alpha}{D} \sum_P |\braket{\psi|P|\psi}|^{2\alpha-1},~~~\text{(Triangle inequality)}
\end{align}
To upper bound the Lipschitz constant, we need to maximize the sum of $|\braket{\psi|P|\psi}|^{2\alpha-1}$ subject to the hypersphere normalization constraint $\sum_P |\braket{\psi|P|\psi}|^{2} = D$. This maximization assumes $|\langle P \rangle| \le 1$ and restricts the replica index to $\alpha > 1.5$ (or integer $\alpha \ge 2$), which ensures the sum is strictly Schur-convex. Intuitively, we wish to push as many expectation values to $\pm 1$ so as to achieve the maximum. Rigorously, using the theory of majorization in Schur-convex functions~\cite{marshall2011inequalities}, we see that taking a stabilizer state $\ket{\psi_{\rm STAB}}$ and restricting the sum over Paulis to the stabilizer group $\mathcal S$  achieves the required maximization
\begin{align}
  \eta \le \frac{4\alpha}{D}\sum_{P \in \mathcal S}|\pm1|^{2\alpha-1} = 4\alpha.
\end{align}
Thus, since the Lipschitz constant is upper bounded by $\bigO{1}$, we can directly utilize L\'evy's lemma as stated in Eq.~\eqref{eq:levy_lemma}.
For Scrooge ensembles with first moment being thermal Gibbs $\sigma_\beta$, we have (for $\beta > 0$)
\begin{align}
 \norm{\sigma_\beta} = e^{-\beta N(\epsilon_0 - f(\beta))},  
\end{align}
where $\epsilon_0$ and $f(\beta)$ are the ground-state energy and free-energy densities, respectively. The operator norm of the thermal density matrix decays exponentially with system size $N$, thus allowing us to use self-averaging for $\xi_\alpha$, and consequently for linearized SRE $M^{\rm lin}_{\alpha}$.

\begin{figure}
    \centering
    \includegraphics[width=0.9\linewidth]{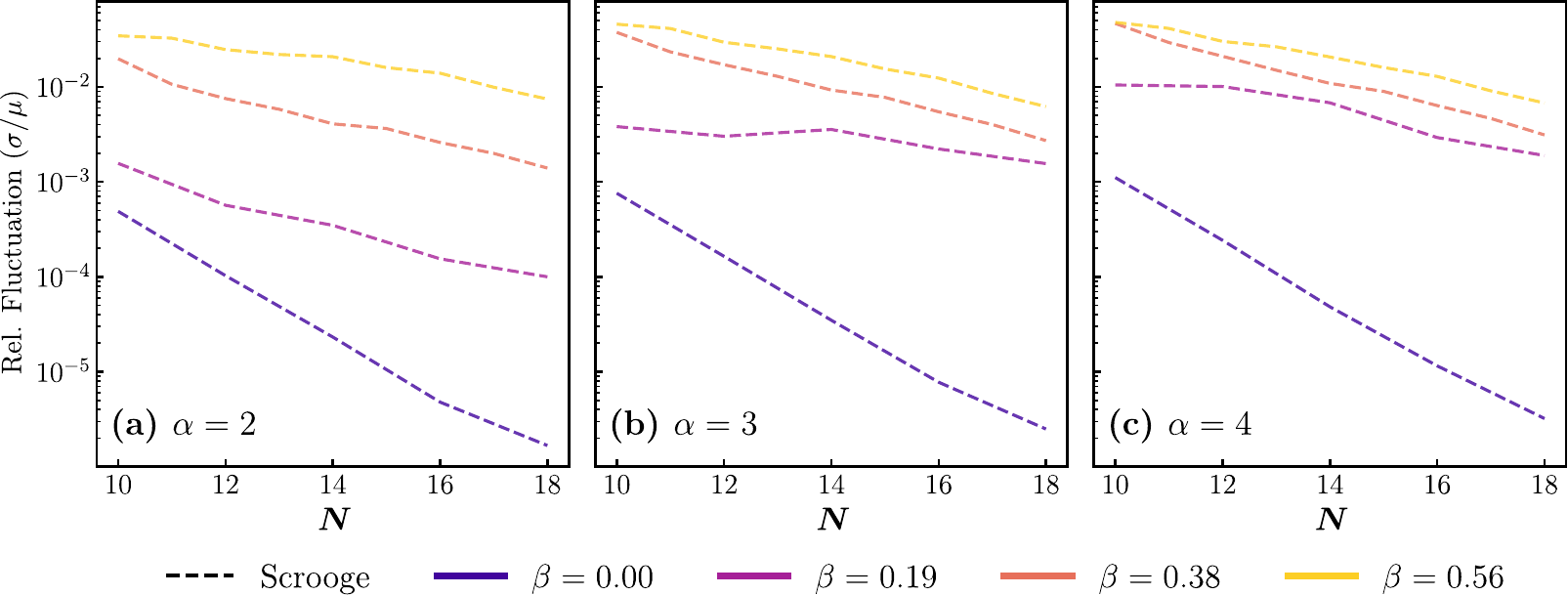}
    \caption{Relative fluctuation of Scrooge fSRE (standard deviation normalized by the mean) vs. system size $N$ for $\beta = 0,0.19,0.38,0.56$. There is an exponential drop in relative fluctuation highlighting concentration around the mean fSRE of the Scrooge ensemble.}
    \label{fig:rel_fluc}
\end{figure}

While we do not prove self-averaging for the actual fSRE, we provide numerical evidence in Fig.~\ref{fig:rel_fluc} where we plot the relative fluctuation of the fSRE (taken as standard deviation $\sigma$ divided by the mean $\mu$) versus system size for various inverse temperatures. We observe that the relative fluctuation decreases exponentially with the system size, with the exponent depending on the inverse temperature $\beta$.

\section{IPR of Scrooge states}
\label{sm:scrooge_magic}

Here we derive Eq.~\eqref{eq:scrooge_ipr_main} of the main text, which gives expressions for the IPR averaged over Scrooge states, first in the regime of fixed nonzero $\beta$ in the thermodynamic limit and then for fixed $N$ as $\beta \to 0$. We consider integer $\alpha$ in this section.

\subsection{Scrooge IPR for $N \to \infty$ and fixed $\beta$}

For every Haar random state $\ket{\phi}$, we approximate the Scrooge-random state as~\cite{mok2026universalityscrooge,mcginley2025scrooge}
\begin{equation}
    \ket{\psi} = \sqrt{D \sigma} \ket{\phi}.
\end{equation}
The $k$th moment of this ensemble is
\begin{equation}
    \tilde{\rho}_\text{Scr.}^{(k)}(\sigma) = \Eset{\phi \sim \text{Haar}(D)} \parens{\ket{\psi}\bra{\psi}}^{\otimes k} = (D \sigma)^{\otimes k} \rho_{\text{Haar}}^{(k)}.
\end{equation}
Comparing to the $k$th moment of the actual Scrooge ensemble, this has an additive error $\bigO{k \norm{\sigma}_2}$ and relative error $\bigO{4^k k \norm{\sigma}_2}$. Indeed, even though $\ket{\psi}$ is, strictly speaking, unnormalized (except in the Haar limit $\sigma \to I/D$), its norm is close to unity,
\begin{equation}
    \E_\psi \sparens{\braket{\psi|\psi}^k} = 1 + \bigO{k^2 \norm{\sigma}_2^2},
\end{equation}
up to a purity correction which is expected to be exponentially small in system size in a typical quantum many-body state~\cite{mok2026universalityscrooge}. Assuming the quenched average is close to the analytically calculated annealed average~\cite{sierant2026theorymagicphasetransitions}, we approximate
\begin{align} \label{eq:self_avg}
 \Eset{\psi \sim \text{Scr.}(\sigma)} (\widetilde M_\alpha(\ket \psi)) \approx \frac{1}{1-\alpha}\log_2\left(\frac{\mathbb E_\psi(\ipr_\alpha)-1}{D-1}\right).
\end{align}
Let $H = \sum_{Q \neq I} a_Q Q$. Using the approximate Scrooge ensemble,
\begin{equation}
    \Eset{\psi \sim \text{Scr.}(\sigma)}\ipr_\alpha(\psi) \approx \sum_{\pi \in S_{2\alpha}} \sum_{P} \Tr((P \sigma)^{\otimes 2\alpha} \hat{\pi}).
\label{eq:scrooge_magic_approx}
\end{equation}
Let us define $t_m(P) = \Tr[(P \sigma)^m]$ and write
\begin{equation}
    \sigma = \frac{1}{D} \sum_P r_P P,
\end{equation}
with $t_1(P) \equiv r_P$. The following inequalities will be useful:
\begin{equation}
    |t_m(P)| \leq \Tr |P \sigma|^m = \Tr(\sigma^m).
\end{equation}
For any $m \geq 2$,
\begin{equation}
    \sum_P |t_1(P)|^m = \sum_P |{r_P}|^m \leq \parens{\sum_P {r_P}^2}^{m/2} = \parens{D \Tr(\sigma^2)}^{m/2}.
\end{equation}
Finally, we note that for generic high-temperature Gibbs states, 
\begin{equation}
    \Tr(\sigma^m) = D^{1-m} \exp\parens{\bigO{N\beta^2}}.
\end{equation}
Now, let us evaluate Eq.~\eqref{eq:scrooge_magic_approx} for $\alpha = 2$. We can write
\begin{equation}
    \Eset{\psi \sim \text{Scr.}(\sigma)}\ipr_2(\psi) \approx \sum_P ({t_1}^4 + 3 {t_2}^2) + R,
\end{equation}

\subsubsection{Bounding $|R|$ for $\alpha = 2$} 

\noindent The remainder term $R$ is bounded by
\begin{equation}
    |R| \leq C \sum_{P} \parens{ |t_2 {t_1}^2| + |t_3 t_1| + |t_4|},
\end{equation}
for some constant $C$ (which comes from the combinatorial factor). Noting that $PQP = f_{PQ} Q$ where $f_{PQ} \in \{\pm 1\}$ is the sign factor,
\begin{equation}
\begin{aligned}
    t_2 &= \Tr(P \sigma P \sigma) \\&= \frac{1}{D^2} \sum_{Q,Q'} \Tr(P Q P Q') r_Q r_{Q'} \\
    &= \frac{1}{D} \sum_{Q} f_{PQ} {r_Q}^2,
\end{aligned}
\end{equation}
and thus
\begin{equation}
    \sum_P t_2(P)^2 = \frac{1}{D^2} \sum_{P,Q,Q^\prime} f_{PQ} f_{PQ^\prime} {r_Q}^2 {r_{Q^\prime}}^2 = \sum_P {t_1}^4 = \sum_P {r_P}^4,
\end{equation}
using the orthogonality relation $\sum_P f_{PQ} f_{PQ^\prime} = D^2 \delta_{QQ^\prime}$ ($f_{PQ}$ is the character of $\mathcal{P}_N$, from which orthogonality holds). This implies that
\begin{equation}
    \Eset{\psi \sim \text{Scr.}(\sigma)}\ipr_2(\psi) \approx 4\sum_P {r_P}^4 + R.
\end{equation}
The first term is four times the IPR of $\sigma$. To bound the remainder term, we use the aforementioned inequalities to get
\begin{equation}
    |R| = \bigO{ D \Tr^2(\sigma^2) + D^{3/2} \Tr(\sigma^3) \Tr^{1/2}(\sigma^2) + D^2 \Tr(\sigma^4)  }.
\end{equation}
Now, substituting the generalized purities of $\sigma$ at high temperature, we get
\begin{equation}
    |R| = \bigO{\frac{1}{D} \exp(\bigO{N \beta^2})}.
\end{equation}
This means that for a sufficiently high constant temperature, $|R|$ is exponentially small in $N$, and we establish
\begin{equation}
    \Eset{\psi \sim \text{Scr.}(\sigma)}\ipr_2(\psi) \approx 4\sum_P {r_P}^4 + \bigO{e^{-cN}} = 4 \ipr_2(\sigma) + \bigO{e^{-cN}}.
\end{equation}
For constant temperature, we expect $\ipr_\alpha(\sigma_\beta)$ to be exponentially large in $N$, so the $-1$ term in the numerator of Eq.~\eqref{eq:self_avg} is negligible, like $|R|$ in the thermodynamic limit. Taking the logarithm, this implies that to leading order in $N$,
\begin{equation}
    M_2 \approx \fsre_2 \approx N - \log_2 (\ipr_2(\sigma_\beta)).
\end{equation}
Using a cluster expansion, we obtain the extensive behavior of the logarithm of the IPR (and subsequently volume-law corrections to the SRE and fSRE),
\begin{equation}
    \ln(\ipr_2(\sigma_\beta)) = {\beta^4 \sum_{Q \neq I} a_Q^4 + \bigO{N\beta^5}},
\end{equation}
so we obtain
\begin{equation} \label{eq:m2_app}
    M_2 \approx \fsre_2 \approx N (1 - c_2\beta^4  + \bigO{\beta^5}),~c_2 = \frac{\sum a_Q^4}{N\ln 2}.
\end{equation}

\subsubsection{Bounding $|R|$ for $\alpha > 2$}

We want to generalize the bounding of the remainder term $R$ from the approximate Scrooge average for arbitrary integer $\alpha > 2$. The approximate Scrooge average involves summing over all permutations $\pi \in S_{2\alpha}$
\begin{align}
\Eset{\psi \sim \text{Scr.}(\sigma)}\ipr_\alpha(\psi) \approx \sum_{\pi \in S_{2\alpha}} \sum_P \text{Tr}((P\sigma)^{\otimes 2\alpha} \hat{\pi}).
\end{align}
Any permutation $\pi \in S_{2\alpha}$ can be decomposed into disjoint cycles. Let the cycle type be characterized by a tuple $\vec{k} = (k_1, k_2, \dots, k_{2\alpha})$, where $k_j$ is the number of cycles of length $j$. The power inside the traces multiplied by the power outside must sum to $2\alpha$, the total length of all cycles. This imposes the strict constraint
\begin{align}
\sum_{j=1}^{2\alpha} j \cdot k_j = 2\alpha.
\end{align}
For a given permutation of cycle type $\vec{k}$, the trace evaluates to
\begin{align}
\text{Tr}((P\sigma)^{\otimes 2\alpha} \hat{\pi}) = \prod_{j=1}^{2\alpha} [t_j(P)]^{k_j},
\end{align}
where $t_j(P) = \text{Tr}((P\sigma)^j)$ and $t_1(P) \equiv r_P \equiv c_P$. Therefore, the total sum is grouped by cycle structures, weighted by a combinatorial factor $C(\vec{k})$ (the number of permutations of that cycle type, which depends \textit{only} on $\alpha$, not $N$)
\begin{align}
\mathbb{E} \ipr_\alpha \approx \sum_{\vec{k} \vdash 2\alpha} C(\vec{k}) \sum_P \prod_{j=1}^{2\alpha} [t_j(P)]^{k_j}.
\end{align}
Note that $\vec{k}$ allows entries to be 0 for non-trivial partitions. Using the triangle inequality and the combinatorial bound, we have
\begin{align}
|R| \le C(\alpha) \sum_{\vec{k} \in R} \sum_P \left| \prod_{j=1}^{2\alpha} t_j(P)^{k_j} \right| = C(\alpha) \sum_{\vec{k} \in R} \sum_P |r_P|^{k_1} \prod_{j=2}^{2\alpha} |t_j(P)|^{k_j}.
\end{align}
We utilize the inequality $|t_j(P)| \le \text{Tr}(\sigma^j)$ for $j \ge 2$. Because this upper bound is independent of $P$, we can pull it out of the sum
\begin{align} \label{eq:remainder_gen}
|R(\vec{k})| \le \left( \prod_{j=2}^{2\alpha} \text{Tr}(\sigma^j)^{k_j} \right) \sum_P |r_P|^{k_1}.
\end{align}
At high temperatures, the generalized purities scale as:
\begin{align}
\text{Tr}(\sigma^j) = D^{1-j} \exp(\mathcal{O}(N\beta^2)).
\end{align}
Let $K' = \sum_{j=2}^{2\alpha} k_j$ be the number of cycles of length $\ge 2$. The product over $j \ge 2$ gives a $D$-scaling of
\begin{align}
D^{\sum_{j=2}^{2\alpha} (1-j)k_j} = D^{K' - \sum_{j=2}^{2\alpha} j k_j} = D^{K' - (2\alpha - k_1)}.
\end{align}
Now, we must evaluate $\sum_P |r_P|^{k_1}$. We split this into three exhaustive cases based on $k_1$. Let $K = k_1 + K'$ be the total number of cycles in the permutation.

\begin{enumerate}
    \item \textbf{Case I: $k_1 \ge 2$.} Using the norm inequality $\sum |x|^{k_1} \le (\sum x^2)^{k_1/2}$ for $k_1 \ge 2$:
\begin{align}
\sum_P |r_P|^{k_1} \le \left(\sum_P r_P^2\right)^{k_1/2} = (D \text{Tr}(\sigma^2))^{k_1/2}.
\end{align}

Since $\text{Tr}(\sigma^2) = D^{-1}\exp(\mathcal{O}(N\beta^2))$, we have $D\text{Tr}(\sigma^2) = \exp(\mathcal{O}(N\beta^2))$.
Therefore, the sum over $P$ contributes {no} $D$ {dependence}, only $\exp(\mathcal{O}(k_1 N\beta^2))$. The overall scaling in $D$ for this term is simply $D^{K' - 2\alpha + k_1} = D^{K - 2\alpha}$.
Because the permutation is not the identity, the number of cycles is strictly $K \le 2\alpha - 1$.
Thus, the term in Eq.~\eqref{eq:remainder_gen} scales at most as $D^{-1} \exp(\mathcal{O}(N\beta^2))$.

\item \textbf{Case II: $k_1 = 0$.} Here, $\sum_P |r_P|^0 = \sum_P 1 = D^2$.
The overall $D$ scaling is $D^2 \times D^{K' - 2\alpha}$. Since $k_1 = 0$, $K' = K$.
We know $\sum_{j=2}^{2\alpha} j k_j = 2\alpha$. Because every cycle has length $\ge 2$, we have $2K \le 2\alpha \implies K \le \alpha$.
The scaling is therefore bounded by:
\begin{align}
D^{2 + K - 2\alpha} \le D^{2 - \alpha}.
\end{align}
For $\alpha = 2$, this exponent is $0$, which is why the $3t_2^2$ term survives at leading order in the preceding $\alpha=2$ analysis. However, for $\alpha \ge 3$, $2-\alpha \le -1$. Thus, for $\alpha \ge 3$, these terms in Eq.~\eqref{eq:remainder_gen} scale at most as $D^{-1} \exp(\mathcal{O}(N\beta^2))$.

\item \textbf{Case III: $k_1 = 1$.} Using Cauchy--Schwarz:
\begin{align}
 \sum_P |r_P| \le \sqrt{D^2 \sum_P r_P^2} = D \sqrt{D \text{Tr}(\sigma^2)} = D \exp(\mathcal{O}(N\beta^2)).
\end{align}
The overall $D$ scaling is $D^1 \times D^{K' - (2\alpha - 1)} = D^{K' - 2\alpha + 2}$.
Since $\sum_{j=2}^{2\alpha} j k_j = 2\alpha - 1$, and each cycle is length $\ge 2$, we have $2K' \le 2\alpha - 1 \implies K' \le \alpha - 1/2$. Because $K'$ is an integer, $K' \le \alpha - 1$.
The scaling is bounded by
\[
D^{(\alpha - 1) - 2\alpha + 2} = D^{1 - \alpha}.
\]

For $\alpha \ge 3$, $1-\alpha \le -2$. These terms scale at most as $D^{-2} \exp(\mathcal{O}(N\beta^2))$ in Eq.~\eqref{eq:remainder_gen} and are even more strongly suppressed.
\end{enumerate}
Thus for any $\alpha \ge 2$, the remainder term scales maximally as $\mathcal{O}(D^{-1} \exp(\mathcal{O}(N\beta^2)))$. Hence it becomes zero in the thermodynamic limit, which allows
\begin{align} \label{eq:scr_ipr_th}
 \Eset{\psi \sim \text{Scr.}(\sigma_\beta)}\ipr_\alpha(\psi) \approx d_\alpha\ipr_\alpha(\sigma_\beta),~d_\alpha\sim \bigO{1},  
\end{align}
with $d_2 = 4,~d_{\alpha \ge 3} = 1$. Thus, since $d_\alpha$ leads to only $\bigO{1}$ differences, we obtain that Scrooge magic density goes to the Gibbs state magic density in the thermodynamic limit for a fixed $\beta$.

\subsection{Scrooge IPR for fixed $N$ and $\beta \to 0$}

There is another term that we have to account for, namely the Haar-like background fluctuations arising from a Gaussian ansatz. For equilibrium states with inverse temperatures close to zero, we treat the expectation values of non-local Paulis $x_P = \langle \psi | P | \psi \rangle$ as normally distributed variables~\cite{turkeshi2025fSRE}. For the non-TRI (time-reversal-symmetry-broken) case, the state coefficients are fully complex, and all nonidentity Pauli observables acquire nonzero fluctuations. The probability density function for each non-local observable $x_P$ is given by the Gaussian measure
\begin{align} \label{eq:gaussian_ansatz}
    P(x_P) = \frac{1}{\sqrt{2\pi b_P}} \exp\left( - \frac{x_P^2}{2b_P} \right),~ b_P = \mathbb E_\psi[\langle \psi | P | \psi \rangle^2] - (\mathbb E_\psi[\langle \psi | P | \psi \rangle])^2.
\end{align}
In the thermodynamic limit, at any nonzero $\beta$ for $\alpha \ge 2$, the contribution from these fluctuating terms is exponentially suppressed for $\alpha>2$ and remains $\bigO{1}$ for $\alpha=2$, while the leading IPR is given as per Eq.~\eqref{eq:scr_ipr_th}, as we will see below. But for finite $N$ systems, for $|\beta| \lesssim \beta_c$ (Eq.~\eqref{eq:beta_c_8}, see below), these terms become the major contributing term arising from sum over non-local Paulis (termed `Gaussian-like' Paulis) where $b_P \approx \sigma \sim 2^{-N}$. To compute the contribution of these non-local Paulis (about $\bigO{D^2}$ such strings) to the IPR, one can use the Gaussian distribution for $\braket{P}_{\psi} \sim \mathcal N(0,2^{-N})$ to get
\begin{align}
  \Eset{\psi \sim \text{Scr.}(\sigma_\beta)}\ipr_\alpha(\psi) \approx\sum_{P,\text{Gaussian}} \braket{P}^{2\alpha}_\psi \approx 1 + \sum_{P \neq I} \int_{-1}^1 dx~\frac{x^{2\alpha}}{\sqrt{2\pi\sigma}}e^{-\frac{x^2}{2\sigma} } \approx 2^{N(2-\alpha)}(2\alpha-1)!! + 1.
\end{align}
Note that the Haar-like term $D^{2-\alpha}(2\alpha-1)!!$ is exponentially suppressed (or constant) for $\alpha > 2$ ($\alpha = 2$) in the thermodynamic limit, while the IPR of the Gibbs state is exponentially large for fixed nonzero $\beta$ at $N \to \infty$. For a fixed $N$ and $\beta \to 0$, $\sigma_\beta \to I/D$ implying $\ipr_\alpha(\sigma_\beta) \to 1$. Thus the actual Scrooge-averaged IPR for $\alpha \ge 2$ becomes
\begin{equation} \label{eq:scrooge_ipr}
    \Eset{\psi \sim \text{Scr.}(\sigma_\beta)}\ipr_\alpha(\psi) \approx
    \begin{cases}
        \displaystyle 2^{N(2-\alpha)} (2\alpha - 1)!! + 1,
        & N \text{ fixed},\ \beta \to 0, \\[0.6em]
        \displaystyle d_\alpha\ipr_\alpha(\sigma_\beta),
        & N \to \infty,\ \beta \text{ fixed}
    \end{cases}
\end{equation}
The crossover inverse temperature $\beta_c$ comes from equating the two IPR expressions in Eq.~\eqref{eq:scrooge_ipr} at a fixed $N$. Taking the thermal-like IPR $\ipr_\alpha(\sigma_\beta) \sim \exp(\bigO{N\beta^{2\alpha}})$ and performing a Taylor expansion at high temperatures (small $\beta$) yields $\sim 1 + N\beta^{2\alpha} + \bigO{N\beta^{2\alpha+1}}$. Ignoring the $\mathcal{O}(1)$ identity term, the leading extensive thermal correction scales as $\sim N\beta^{2\alpha}$. Equating this thermal contribution to the Haar-like Gaussian fluctuation term isolates the crossover regime
\begin{align} \label{eq:beta_c_8}
    d_\alpha N\beta_c^{2\alpha} \approx D^{2-\alpha}\frac{(2\alpha)!}{ 2^\alpha \alpha!} \implies \beta_c \approx \left(\frac{(2\alpha-1)!!}{d_\alpha}\right)^{\frac{1}{2\alpha}}N^{-\frac{1}{2\alpha}}2^{-N\frac{\alpha-2}{2\alpha}}.
\end{align}
More accurately, using Eq.~\eqref{eq:gibbsipr_cluster} (in Sec.~\ref{sm:gibbs_ipr} later) gives us
\begin{align} \label{eq:beta_c_8_accurate} d_\alpha\left(\sum_Qa_Q^{2\alpha}\right)\beta_c^{2\alpha} \approx D^{2-\alpha}\frac{(2\alpha)!}{2^\alpha \alpha!} \implies \beta_c \approx \left(\frac{(2\alpha-1)!!}{d_\alpha c_\alpha \ln2}\right)^{\frac{1}{2\alpha}}N^{-\frac{1}{2\alpha}}2^{-N\frac{\alpha-2}{2\alpha}}.
\end{align}
For calculating magic densities in the thermodynamic limit, $\bigO 1$ constants can be ignored, and we propose the interpolating formula for the Scrooge-averaged IPR as
\begin{align} \label{eq:ipr_partition_scr}
 \Eset{\psi \sim \text{Scr.}(\sigma_\beta)}\ipr_\alpha(\psi) \approx 2^{N(2-\alpha)}(2\alpha-1)!!+d_\alpha\ipr_\alpha(\sigma_\beta),
\end{align}
which we use in Fig.~\ref{fig:magic_density} of the main text. Note that this formula, strictly speaking, does not yield the correct limit for $\alpha = 2$ and $\beta \to 0$, where the Haar-like background term is of the same order as the thermal contribution. However, this discrepancy is only $\bigO 1$ and does not affect the leading order behavior of the magic density in the thermodynamic limit.

\section{Approximate Scrooge variance}
To evaluate the high-temperature expansions and the emergent Haar-like background (from the Gaussian ansatz described at the end of Sec.~\ref{sm:scrooge_magic}), we analyze the variance of the Pauli observables, $b_P$ (Eq.~\eqref{eq:gaussian_ansatz}), under the approximate Scrooge ensemble first, and then under the exact Scrooge ensemble later.

\noindent We wish to evaluate the contribution of the non-local ``Gaussian-like'' Paulis to the magic, so we integrate the Gaussian measure described in Eq.~\eqref{eq:gaussian_ansatz} to evaluate the $2\alpha$-th moment for integer $\alpha \geq 0$:
\begin{align} \label{eq:gaussian_integral}
    \mathbb{E}_{\psi}[x_P^{2\alpha}] &= \int_{-\infty}^{\infty} x_P^{2\alpha} P(x_P) dx_P \nonumber \\
    &= b_P^\alpha (2\alpha-1)!! \nonumber \\
    &= b_P^\alpha \frac{(2\alpha)!}{2^\alpha \alpha!}.
\end{align}
For noninteger $\alpha$, one can use the analytically continued formula for the random variable $|X| = |x_P|$:
\begin{align} \label{eq:gaussian_integral_nonint}
    \mathbb{E}_{\psi}[|x_P|^{2\alpha}] 
    &= b_P^\alpha 2^\alpha \frac{\Gamma(\alpha + 1/2)}{\sqrt{\pi}}.
\end{align}
The rest of this section and the following sections assume that $\alpha$ is a positive integer, with appropriate analytically continued formulas for noninteger $\alpha$ highlighted as required. Under the Gaussian ansatz, summing Eq.~\eqref{eq:gaussian_integral} over the $D^2-1$ nonidentity Pauli strings gives the Haar-like background contribution
\begin{align} \label{eq:haar_background_sum}
    \mathbb{E}_{\psi}[\ipr_\alpha(\psi)]\approx\sum_{P, \text{Gaussian}} \mathbb{E}_{\psi}[x_P^{2\alpha}] \approx \frac{(2\alpha)!}{2^\alpha \alpha!} \sum_{P \ne I} b_P^\alpha.
\end{align}
We therefore evaluate this contribution by computing the Scrooge variances $b_P$.

We compute the variance using the approximate (unnormalized) Scrooge ensemble. The $k$-th moment of this simplified ensemble is given by $\tilde{\rho}^{(k)}_{Scr} = \mathbb{E}_{\phi \sim \text{Haar}} [ (|\tilde{\psi}\rangle\langle\tilde{\psi}|)^{\otimes k} ] = (D\sigma_\beta)^{\otimes k} \rho^{(k)}_{\text{Haar}}$, where $|\tilde{\psi}\rangle = \sqrt{D\sigma_\beta}|\phi\rangle$ is the unnormalized state.
For the variance ($k=2$), we use the Haar 2-design
\[\rho^{(2)}_{\text{Haar}} = \frac{\mathcal{I} + \mathcal{F}}{D(D+1)},\]
where $\mathcal I$ is the identity permutation and $\mathcal F$ is the flip permutation. The approximate variance $\tilde{b}_P$ reads:
\begin{align}
    \tilde{b}_P &= \Tr\left( (P \otimes P) \tilde{\rho}^{(2)}_{Scr} \right) - (\Tr(P\sigma_\beta))^2 = \frac{D^2}{D(D+1)} \Tr\left( (P \sigma_\beta \otimes P \sigma_\beta)(\mathcal{I} + \mathcal{F}) \right)- (\Tr(P\sigma_\beta))^2 \notag \\
    &= \frac{1}{D+1} \left( D\Tr(P\sigma_\beta P\sigma_\beta) - (\Tr(P\sigma_\beta))^2 \right) \notag \\ \label{eq:appscr_var}
    &\approx \Tr(P\sigma_\beta P\sigma_\beta).
\end{align}
We will use this variance in calculating asymptotic behavior of magic in the next section.

\section{High-temperature expansion of the Gibbs IPR}

\label{sm:gibbs_ipr}

We now compute the power-series expansion of the inverse participation ratio ($\ipr_\alpha$) of the thermal Gibbs state $\sigma_\beta$ in terms of the inverse temperature $\beta$. Note that this derivation holds for integer $\alpha \ge 2$. For real $\alpha$, one can take the current formulae as the analytic continuation of the integer-valued $\alpha$ ones.  We define the normalized Hamiltonian moments projected onto a Pauli operator $P$ as:
\begin{align} \label{eq:nu_k_def_13}
    \nu_{k}(P) = \frac{\Tr(P H^{k})}{D}
\end{align}
where $D = 2^N$ is the Hilbert space dimension. Without loss of generality, we shift the energy scale such that the Hamiltonian is traceless (already the case for our MFIM Hamiltonian taken in the main text), $\Tr(H) = 0$, which implies $\nu_1(I) = 0$.

The methodology begins by expanding the thermal Gibbs state $\sigma_\beta = e^{-\beta H}/Z$ order-by-order in $\beta$. For a non-identity Pauli string ($P \ne I$), the unnormalized expectation value follows from the standard Taylor series of the exponential operator
\begin{align*}
    &\Tr(P e^{-\beta H}) = \sum_{k=0}^{\infty} \frac{(-\beta)^k}{k!} \Tr(P H^k) \\
    &= D \left( -\beta \nu_1(P) + \frac{\beta^2}{2} \nu_2(P) - \frac{\beta^3}{6} \nu_3(P) + \mathcal{O}(\beta^4) \right).
\end{align*}
Notice that the $k=0$ term vanishes strictly because $\Tr(P) = 0$ for non-identity Paulis. Concurrently, the partition function evaluates to $Z = \Tr(e^{-\beta H}) = D(1 + \frac{\beta^2}{2}\nu_2(I) + \mathcal{O}(\beta^3))$. By dividing the unnormalized expectation by $Z$, the exact Pauli coefficients $c_P(\beta) = \Tr(P\sigma_\beta)$ take the following form up to second order
\begin{align} \label{eq:cp_expansion_14}
    c_P(\beta) &= \frac{-\beta \nu_1(P) + \frac{\beta^2}{2} \nu_2(P) + \mathcal{O}(\beta^3)}{1 + \frac{\beta^2}{2}\nu_2(I) + \mathcal{O}(\beta^3)} \nonumber \\
    &= -\beta \nu_1(P) + \frac{\beta^2}{2} \nu_2(P) + \mathcal{O}(\beta^3).
\end{align}
To evaluate the thermal $\ipr$, we must raise these state coefficients to the $2\alpha$-th power. We evaluate this moment by applying the binomial expansion to Eq.~\eqref{eq:cp_expansion_14}. By factoring out the leading $\beta$ dependence, we analytically calculate the primary $\beta^{2\alpha}$ coefficient and its first sub-leading correction
\begin{align} \label{eq:cp_2alpha_expansion_15}
    c_P(\beta)^{2\alpha} &= \left( -\beta \nu_1(P) + \frac{\beta^2}{2} \nu_2(P) + \mathcal{O}(\beta^3) \right)^{2\alpha} \nonumber \\
    &= (-\beta \nu_1(P))^{2\alpha} + \binom{2\alpha}{1} (-\beta \nu_1(P))^{2\alpha-1} \left( \frac{\beta^2}{2} \nu_2(P) \right) + \mathcal{O}(\beta^{2\alpha+2}) \nonumber \\
    &= \beta^{2\alpha} \nu_1(P)^{2\alpha} - 2\alpha \beta^{2\alpha-1} \nu_1(P)^{2\alpha-1} \left( \frac{\beta^2}{2} \nu_2(P) \right) + \mathcal{O}(\beta^{2\alpha+2}) \nonumber \\
    &= \nu_{1}(P)^{2\alpha}\beta^{2\alpha} - \alpha\nu_{1}(P)^{2\alpha-1}\nu_{2}(P)\beta^{2\alpha+1} + \mathcal{O}(\beta^{2\alpha+2}).
\end{align}   
Finally, the full thermal IPR $\ipr_\alpha(\sigma_\beta)$ is obtained by summing $c_P(\beta)^{2\alpha}$ over all Pauli operators. Recognizing that the identity Pauli yields a constant ($c_I(\beta)^{2\alpha} = 1$), we explicitly separate it from the summation. Substituting the analytical expansion from Eq.~\eqref{eq:cp_2alpha_expansion_15} back into the sum over all $P \ne I$ yields our exact perturbative formula
\begin{align} \label{eq:f_alpha_beta_expansion_16}
    \ipr_{\alpha}(\sigma_\beta) &= 1 + \sum_{P \ne I} c_P(\beta)^{2\alpha} \nonumber \\
    &= 1 + \sum_{P \ne I} \bigg( \nu_{1}(P)^{2\alpha}\beta^{2\alpha} - \alpha\nu_{1}(P)^{2\alpha-1}\nu_{2}(P)\beta^{2\alpha+1} + \mathcal{O}(\beta^{2\alpha+2}) \bigg).
\end{align}
For $H = \sum a_Q Q$ with $\Tr(H) = 0$, there is no $Q = I$ term. Thus,
\begin{align}
 \sum_{P \ne I}\nu_{1}(P)^{2\alpha}  = \sum_Q a_Q^{2\alpha} \propto N.
\end{align}
For non-integer $\alpha$, one should use $|a_Q|^{2\alpha}$. This finite-$N$ Taylor expansion yields a leading nontrivial term proportional to $N\beta^{2\alpha}$ in Eq.~\eqref{eq:f_alpha_beta_expansion_16}. However, taking the thermodynamic limit ($N \to \infty$) at a fixed temperature $\beta > 0$ causes this polynomial truncation to diverge. To properly transition to the macroscopic regime, we must invoke two standard assumptions. First, we assume a finite radius of convergence, restricting our analysis to the strict high-temperature regime (where the local interaction strength satisfies $\beta J \ll 1$) so that the perturbative series is initially well-behaved. Second, we assume the clustering of correlations, namely that thermal correlations between spatially separated local Pauli operators decay rapidly at high temperatures. Under these conditions, the linked-cluster theorem gives us
\begin{align} \label{eq:gibbsipr_cluster}
   \ln\ipr_{\alpha}(\sigma_\beta) =  \beta^{2\alpha} \sum_{Q \neq I} |a_Q|^{2\alpha} + \bigO{N\beta^{2\alpha + 1}}\sim \bigO{N\beta^{2\alpha}}.
\end{align}

\section{Asymptotic behavior of magic}
\label{sm:asymp_magic}
We now evaluate the asymptotic expressions for fSRE and SRE when $N$ is large. We take the generalized inverse participation ratio (IPR) over the full Pauli group $\mathcal{P}_N$ for state $\rho = \ket{\psi}\bra{\psi}$, where $\psi \sim \text{Scrooge}(\sigma_\beta)$ (or $\psi \sim \text{Approx. Scrooge}(\sigma_\beta)$) as:
\begin{align} \label{eq:ipr_full_def}
    \ipr_{\alpha}({\psi}) = \sum_{P \in \mathcal{P}_N} \langle \psi | P | \psi \rangle^{2\alpha} \approx D^{2-\alpha}\frac{(2\alpha)!}{2^\alpha \alpha!} + d_\alpha\ipr_\alpha(\sigma_\beta),
\end{align}
where we have used Eq.~\eqref{eq:ipr_partition_scr} in the last approximate equality. One also notes that due to numerical concentration demonstrated in Fig.~\ref{fig:rel_fluc}, we work with annealed averages in SRE ($M_\alpha(\psi)$) and fSRE ($\fsre_\alpha(\psi)$). Hence, we only need to consider Eq.~\eqref{eq:ipr_full_def}. We take $M_\alpha \equiv M_\alpha (\psi),~\fsre_\alpha \equiv \fsre_{\alpha}({\psi})$ and $\ipr_\alpha \equiv \ipr_{\alpha}({\psi})$ in the following subsections.

\subsection{Regime I: $\alpha \ge 2$}
There are distinct regimes to consider. At any fixed nonzero $\beta$ in the thermodynamic limit, the generalized purities are exponentially large in $N$, which puts $\ipr_\alpha \gg 1$. To see this, we analyze Eq.~\eqref{eq:ipr_full_def} in the thermodynamic limit. For $\alpha \ge 2$, the Haar-like non-local fluctuation term scales as $D^{2-\alpha} = 2^{N(2-\alpha)}$, which is $\mathcal{O}(1)$ for $\alpha = 2$ and exponentially decaying for $\alpha > 2$. Conversely, the thermal term $\ipr_\alpha(\sigma_\beta)$ scales exponentially with $N$ at any finite temperature in the thermodynamic limit (see Eq.~\eqref{eq:gibbsipr_cluster}). Therefore, the thermal contribution dominates entirely
\begin{align} \label{eq:ipr_alpha_ge_2}
    \ipr_\alpha \gg 1.
\end{align}
This also roughly holds for finite $N$ systems with $|\beta| > \beta_c$. In this regime, we have
\begin{align} \label{eq:regime_I}
    \fsre_{\alpha} &= \frac{N - \log_{2}\ipr_{\alpha}}{\alpha-1} + \bigO{2^{-N},\ipr^{-1}_\alpha} \notag \\
    M_\alpha &= \frac{N - \log_{2}\ipr_{\alpha}}{\alpha-1}
\end{align}
For finite N systems, we can have $\ipr_\alpha\sim \bigO{1}$ for $|\beta| \sim \beta_c$ (Eq.~\eqref{eq:beta_c_8}) which then leads to
\begin{align}
  M_\alpha \approx \fsre_\alpha = \frac{N}{\alpha-1} + \bigO{1}.
\end{align}
For $|\beta| < \beta_c$, $\ipr_\alpha-1\ll 1$, which leads to
\begin{align}
   \fsre_\alpha &= \frac{N - \log_2(\ipr_\alpha-1)}{\alpha-1} + \bigO{2^{-N}}, \notag \\
   M_\alpha &= \frac{N}{\alpha-1} + \bigO{\ipr_\alpha-1}.
\end{align}
$\ipr_\alpha$ also becomes smaller than the Haar-like fluctuation of $\bigO{D^{2-\alpha}}$. For these values of $|\beta|$, the accurate form is given by Taylor expanding around $\beta = 0$. For $\alpha > 2$, since $\beta_c \to 0$ as $N \to \infty$, this becomes a singularity in the magic density at $\beta = 0$. In the next subsection, we analyze the Taylor expansion for $\alpha < 2$.

\subsection{Regime II: $\alpha < 2$}
\label{sm:asymp_2}

Note that the Taylor expansion around $\beta = 0$ is only valid for the condition $|\beta| \lesssim \beta_c$ for $\alpha \ge 2$. But, this condition does not apply anymore as $\beta_c \to \infty$ for $\alpha <2$ in the thermodynamic limit, allowing the Taylor expansion to be accurate for larger $\beta$ (dependent upon the order to which we expand). Since we are expanding about $\beta = 0$ here, we utilize the approximate Scrooge formalism (see Sec.~\ref{sm:scrooge_magic}). Taking the variance of $\braket{P}_\psi$ as per Eq.~\eqref{eq:appscr_var} and expanding $\sigma_\beta$ in powers of $\beta$ gives us
\begin{align}
 \tilde b_P &= \frac{1}{D}\bigg[1 + \beta^2 \frac{\Tr(PHPH)}{D} -\beta^3 \frac{\Tr(PHPH^2)}{D} + \beta^4\bigg(\frac{\Tr(PHPH^3)}{3D} + \frac{\Tr(PH^2PH^2)}{4D}-\frac{\Tr(PHPH)\Tr(H^2)}{D^2} \notag \\
 & - \frac{(\Tr(H^2))^2}{4D^2}\bigg)\bigg].
\end{align}
Let $\tilde \epsilon_P$ be the relative finite-temperature perturbation such that $\tilde b_P = \frac{1}{D}(1 + \tilde \epsilon_P)$. Let us also take $H = \sum_Q a_QQ $ as the expansion of the Hamiltonian in the Pauli basis. We carefully track the magnitudes of the terms comprising $\tilde \epsilon_P$:

\begin{itemize}
    \item $\Tr(PHPH) = \sum_{Q_1,Q_2}a_{Q_1}a_{Q_2}\tr(PQ_1PQ_2) = D\sum_Qa^2_Qf_{PQ}\equiv DE_P$.

    \item Let us take $H^2 = \sum_{Q_1,Q_2}a_{Q_1}a_{Q_2}Q_1Q_2 = \sum_{Q}b_QQ$. Then we have
    \begin{align}
        \tr(PHPH^2) = \sum_{Q_1,Q_2}a_{Q_1}b_{Q_2}\tr(PQ_1PQ_2) = D\sum_{\bar Q}a_{\bar Q}b_{\bar Q}f_{P\bar{Q}} \equiv DJ_P.
    \end{align}
    Here $\bar Q$ represents the Pauli strings which are present in both $H$ and $H^2$ (about $\bigO{N}$ such strings). We also have
    \begin{align}
        \tr(PH^2PH^2) = \sum_{Q_1,Q_2}b_{Q_1}b_{Q_2}\tr(PQ_1PQ_2) = D\sum_{ Q}b^2_{Q}f_{P{Q}} \equiv D\tilde E_P.
    \end{align}
    The final sum here is over Pauli strings $Q$ in $H^2$ (about $\bigO{N^2}$ such strings).

    \item Let us take $H^3 = \sum_{Q_1,Q_2,Q_3}a_{Q_1}a_{Q_2}a_{Q_3}Q_1Q_2Q_3 = \sum_{Q}m_QQ$. Then we have
    \begin{align}
        \tr(PHPH^3) = \sum_{Q_1,Q_2}a_{Q_1}m_{Q_2}\tr(PQ_1PQ_2) = D\sum_{ Q'}a_{Q'}m_{ Q'}f_{P{Q'}} \equiv DM_P.
    \end{align}
    Here $Q'$ represents the Pauli strings which are present in both $H$ and $H^3$ (about $\bigO{N}$ such strings).

    \item $\Tr(H^2) = D\sum_Qa^2_Q \equiv Dk$. Notice that the coefficient of identity Pauli in $H^2$, $b_I = k$.%
\end{itemize}
Notice that $\sum_{P}E_P = \sum_{P}J_P  = \sum_{P} M_P = 0$ as they all involve the sum $\sum_{P} f_{PQ}$ for non-identity string $Q$ which evaluates to $0$, as $\sum_P f_{PQ} = D^2\delta_{Q,I}$. The sum $\sum_{P}\tilde E_P$ is trickier as there is the identity string $I$ in $H^2$. Hence, noticing that $b_I = k$ in $H^2$, we have
\begin{align}
    \sum_{P} \tilde E_P = \sum_Qb_Q^2\sum_P f_{PQ}  = D^2b_I^2=D^2k^2.
\end{align}
We now do a Taylor expansion to obtain terms up to $\bigO{\beta^4}$. This requires us to do a second-order Taylor expansion (as coefficient of $\beta^4$ can get a contribution from the second-order squared coefficient of $\beta^2$ term). Thus we have
\begin{align}
    \sum_{P} \tilde b^\alpha_P &= \frac{1}{D^\alpha}\sum_{P}(1 + \tilde \epsilon_P)^\alpha \notag \\
    &\approx \frac{1}{D^\alpha}\sum_{P}\left(1 + \binom{\alpha}{1}\tilde \epsilon_P + \binom{\alpha}{2}\tilde \epsilon^2_P\right).
\end{align}
Using our previously mentioned results over sum of Paulis, and the fact that
\begin{align}
 \sum_{P} E_P^2 &= \sum_{P} \sum_{Q, Q'} a_Q^2 a_{Q'}^2 f_{PQ} f_{PQ'} \nonumber \\
    &= D^2 \sum_Q a_Q^4,   
\end{align}
where we successfully utilized the orthogonality of Pauli characters: $\sum_{P} f_{PQ} f_{PQ'} = D^2 \delta_{Q, Q'}$. We also have
\begin{align}
 \sum_{P}  \left({\tilde E_P} -\left(\frac{\Tr(H^2)}{D}\right)^2\right) = D^2(k^2-k^2) = 0. 
\end{align}
Combining these results, we have that
\begin{align} \label{eq:fin_appvaralpha}
    \sum_{P} \tilde b^\alpha_P \approx D^{2-\alpha}\left[1 + \binom{\alpha}{2}\beta^4 \sum_Qa_Q^4\right].
\end{align}
Now, we need to exclude the contribution from the identity Pauli string $I$ to get $\sum_{P \neq I} \tilde b^\alpha_P$. But since the contribution is of subleading order $\bigO{D^{-\alpha}}$, we can omit it in the thermodynamic limit as compared to the $\bigO{D^{2-\alpha}}$ scaling in Eq.~\eqref{eq:fin_appvaralpha}. Applying this to Eq.~\eqref{eq:haar_background_sum}, we arrive at
\begin{align} \label{eq:scrooge_haar_dip}
    \fsre_{\alpha} \approx N\left(1 - \frac{\alpha c_2}{2}\beta^4 + \mathcal{O}(\beta^5)\right) - \frac{1}{\alpha-1}\log_2\left(\frac{(2\alpha)!}{2^\alpha \alpha!}\right),
\end{align}
where $c_2 \sim \bigO{1}$ as in Eq.~\eqref{eq:m2_app}.

\section{Exact Scrooge analysis}

We now derive the corresponding results using the exact Scrooge measure.

\subsection{Exact Scrooge variance up to $\mathcal{O}(\beta^2)$}
For $\ket{\psi}$ drawn from Scrooge$(\sigma_\beta)$, and any measurable function $f$, the Scrooge average can be written as
\begin{equation}
    \Eset{\psi \sim \text{Scrooge}(\sigma_\beta)}\sparens{f(\psi)} = D \E_{\phi} \sparens{\braket{\phi|\sigma_\beta|\phi} f\parens{\frac{\sqrt{\sigma_\beta}\ket{\phi}}{\braket{\phi|\sigma_\beta|\phi}^{1/2}}}},
\end{equation}
where $\E_\phi$ denotes the expectation over Haar random states. Evaluating this for variance $b_P$ (see~\eqref{eq:gaussian_ansatz}) with $P \neq I$, we obtain
\begin{align}
\label{eq:bp_exact_scrooge}
    b_P = \mathbb{E}_{\phi} \left[ \frac{\Tr^2(P \sqrt{D\sigma_\beta} |\phi\rangle\langle\phi| \sqrt{D\sigma_\beta})}{Z_\phi} \right] - \sparens{\Tr(P\sigma_\beta)}^2,
\end{align}
with $Z_\phi = \langle \phi | D\sigma_\beta | \phi \rangle$.
Let $\Pi_\phi = |\phi\rangle\langle\phi|$. We expand the unnormalized state up to $\mathcal{O}(\beta^2)$,
\begin{align}
    \sqrt{D\sigma_\beta} \Pi_\phi \sqrt{D\sigma_\beta} &= \Pi_\phi - \frac{\beta}{2}(\Pi_\phi H + H \Pi_\phi) + \frac{\beta^2}{8}\left(\Pi_\phi H^2 + H^2 \Pi_\phi + 2 H \Pi_\phi H - \frac{4\Tr(H^2)}{D}\Pi_\phi\right) + \mathcal{O}(\beta^3).
\end{align}
The normalization factor has the expansion $Z_\phi = 1 - \beta \Tr(\Pi_\phi H) + \frac{\beta^2}{2}\left(\Tr(\Pi_\phi H^2) - \Tr(H^2)/D\right) + \bigO{\beta^3}$. %
Evaluating the Haar integral in Eq.~\eqref{eq:bp_exact_scrooge}, we obtain $b_P$ for Scrooge$(\sigma_\beta)$ through $\bigO{\beta^2}$:
\begin{align} \label{eq:exact_variance}
    b_P = \frac{1}{D+1} - \frac{\beta^2}{D(D+3)} \left( \frac{\Tr(H^2)}{(D+1)} - {\Tr(PHPH)} + \frac{3\Tr(PH)^2}{D} \right) + \bigO{\beta^3}.
\end{align}

\subsection{fSRE for $\alpha < 2$}

Expanding the Hamiltonian in the Pauli basis as $H = \sum_Q a_Q Q$, we identify the essential contributions to the variance:
\begin{enumerate}
    \item $\Tr(H^2) = D \sum_Q a_Q^2 = \bigO{DN}$
    \item $\Tr(PHPH) = D \sum_Q a_Q^2 f_{PQ} \equiv D \cdot E_P$
    \item $\Tr(PH)^2 = D^2 a_P^2$
\end{enumerate}
where $f_{PQ} \in \{1, -1\}$ is the Pauli commutation character. Substituting these trace identities into the variance formula, and approximating $D+1 \approx D$ and $D(D+3) \approx D^2$ for large $D$, we get
\begin{align}
    b_P &= \frac{1}{D} - \frac{\beta^2}{D^2} \left[ \sum_Q a_Q^2 - \Tr(PHPH) + \frac{3 \Tr(PH)^2}{D} \right] \nonumber \\
    &= \frac{1}{D} \left[ 1 + \bigO{\frac{\beta^2 N}{D}} + \beta^2 E_P - 3 \beta^2 a_P^2 \right].
\end{align}

Let $\epsilon_P$ be the relative finite-temperature perturbation such that $b_P = \frac{1}{D}(1 + \epsilon_P)$. We carefully track the magnitudes of the terms comprising $\epsilon_P$:
\begin{itemize}
    \item $\bigO{\frac{\beta^2 N}{D}}$: A uniform shift that strictly vanishes in the thermodynamic limit.
    \item $3 \beta^2 a_P^2$: An $\mathcal{O}(1)$ correction, but one that is exclusively nonzero for the $\bigO{N}$ local Pauli terms present in the Hamiltonian.
    \item $\beta^2 E_P$: This term is the same as that described in Sec.~\ref{sm:asymp_2}. Crucially, it is nonzero for almost all $D^2$ Pauli operators.
\end{itemize}

Since the dominant term is $E_P$, which leads to the $\bigO{\beta^4}$ correction in the previous Sec.~\ref{sm:asymp_2} (when evaluated to second order in $\epsilon_P$), we obtain the same fSRE formula as in Eq.~\eqref{eq:scrooge_haar_dip}. Since calculating the explicit $\bigO{\beta^4}$ term in $b_P$ is difficult analytically, we highlight this formula as approximate (deriving it from exact Scrooge), supported by numerical checks.

\section{Supplementary numerical results}
\label{sm:supp_numerics}

We give a brief overview of additional numerical results for other physical setups.

\subsection{Bipartite entanglement}

\begin{figure}
    \centering
    \includegraphics[width=0.9\linewidth]{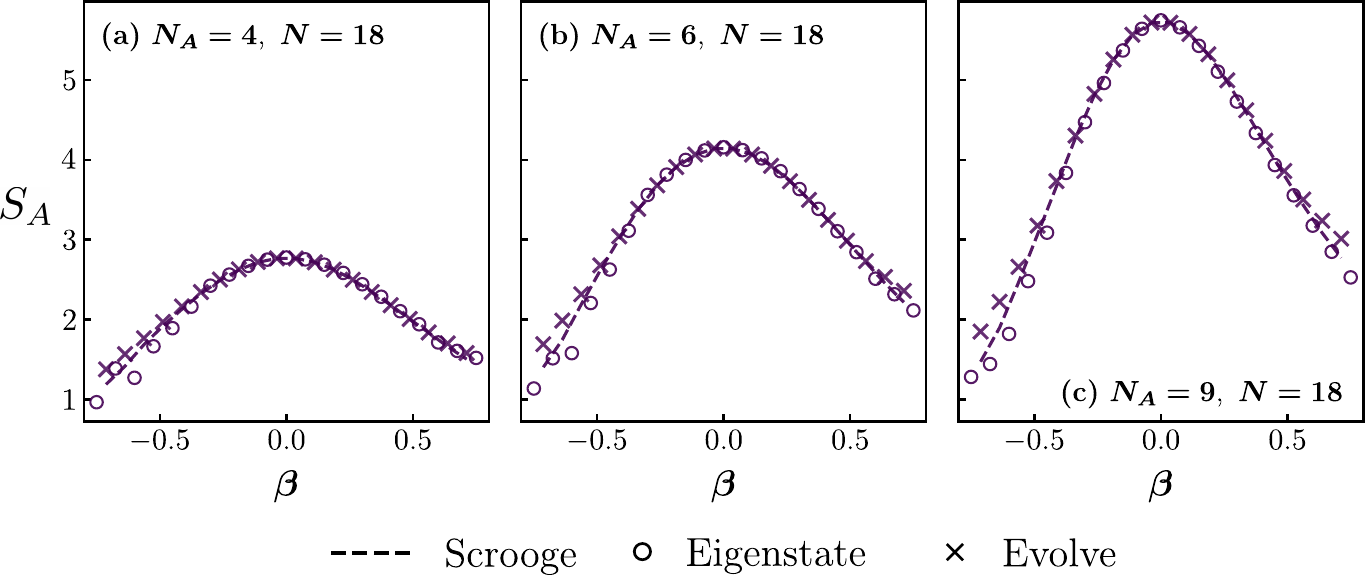}
    \caption{von Neumann entanglement entropy versus inverse temperature $\beta$ for an $N=18$ MFIM setup with various subsystem sizes: (a) $N_A = 4$, (b) $N_A = 6$, and (c) $N_A = 9$. Curves are compared for eigenstates, Scrooge states, and late-time-evolved random product states.}
    \label{fig:supp_ent}
\end{figure}

It has been shown in Refs.~\cite{rodriguez2024eeinfinitetemp,langlett2025eefinitetemp} how the von Neumann bipartite entanglement entropy of finite- and infinite-temperature eigenstates of a chaotic Hamiltonian can be captured by states drawn from the Bianchi--Don\`a (BD) distribution~\cite{bianchi2019bddist}. A priori, it is not clear whether the bipartite entanglement entropy of a Scrooge ensemble at inverse temperature $\beta$ would match that of the corresponding chaotic eigenstate after accounting for the corrections obtained via the BD distribution. Here, we show that the Scrooge ensemble is able to reproduce the bipartite entanglement entropy of chaotic eigenstates at finite temperature as well (see also~\cite{nakagawa2018universality} for a related observation).

Figure~\ref{fig:supp_ent} shows the entanglement entropy $S_A = -\Tr(\rho_A\ln(\rho_A))$, where $\rho_A$ is the reduced density matrix on the first $N_A$ sites of the $N$-qubit system. The entanglement entropy for late-time evolved states agrees well with the average entanglement entropy of Scrooge($\sigma_\beta$), without any corrections. For the eigenstate data plotted in Fig.~\ref{fig:supp_ent}, we apply the corrections from the BD distribution~\cite{langlett2025eefinitetemp},
\begin{align}
  S_A^{\rm eig} = S_A - \frac{f + \log(1-f)}{2} + \sqrt{\frac{n(1-n)}{2\pi}}\left|\log\left(\frac{1-n}{n}\right)\right|\delta_{f,1/2}\sqrt{N}  
\end{align}
where $f = N_A/N$ is the subsystem fraction and $n$ is defined by
\begin{align}
    n = \frac{1}{2}\left(\frac{E_\beta}{E^*} + 1\right),
\end{align}
where $E^*$ is $N$ times the root-mean-square of the Pauli coefficients in the Hamiltonian, which in this case is the MFIM
\begin{align}
  H_{\rm MFIM} = \sum_{i=1}^N g_i X_i + \sum_{i=1}^N h_iZ_i + J\sum_{i=1}^{N-1}Z_iZ_{i+1} + H_1,
\end{align}
where $H_1 = Y_{N-2}Z_{N-1} + Y_{N-1}Z_N$ is added to break time-reversal symmetry.

\subsection{Quantum many-body scars}

\begin{figure}
    \centering
    \includegraphics[width=\linewidth]{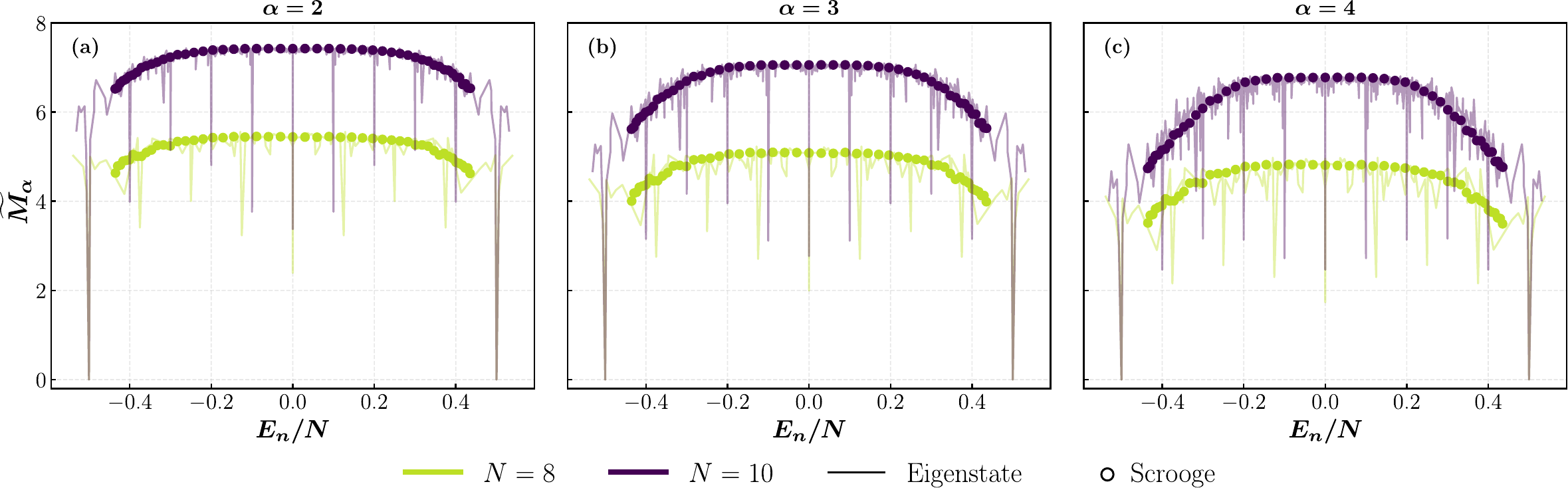}
    \caption{Magic (fSRE) against energy density $E_n/N$ for the Hamiltonian in Eq.~\eqref{eq:scar_ham} (with $\Omega = 1$) showing scar states violating ETH. The magic of eigenstates exhibits sharp dips at periodic intervals, while Scrooge state magic captures the behavior of typical eigenstates but not the atypical scar states.}
    \label{fig:supp_scars}
\end{figure}

Consider the following Hamiltonian with periodic boundary conditions, introduced in Ref.~\cite{choi2019emergent}:
\begin{align} \label{eq:scar_ham}
    H_{\rm Scar} = \frac{\Omega}{2}\sum_{i=1}^N X_i + \sum_{i=1}^N P_{i,i+1}Z_{i+2},\quad P_{i,i+1} = (1-\vec{\sigma}_i\cdot\vec{\sigma}_{i+1})/4.
\end{align}
This Hamiltonian hosts quantum many-body scar states~\cite{choi2019emergent}, namely atypical eigenstates that violate ETH. This behavior manifests itself as dips in the half-chain entanglement entropy~\cite{turner2018weak}, but how does it affect magic?

Figure~\ref{fig:supp_scars} illustrates the filtered stabilizer Rényi entropy (fSRE), $\widetilde{M}_\alpha$ (i.e., our measure of nonstabilizerness or magic) as a function of the energy density for the eigenstates of the Hamiltonian defined in Eq.~\eqref{eq:scar_ham}. Strikingly, the eigenstate magic exhibits sharp, periodic reductions at equispaced energy intervals of $E_n = \Omega n$, where $n \in \{-N/2, -N/2 + 1, \dots, N/2\}$. These deviations signify the presence of quantum many-body scar states. The dips in the magic of these atypical scar states are consistent with Ref.~\cite{hartse2025stabscars}, which conjectures that their magic scales as $\bigO{\log N}$, much smaller than the volume-law scaling of typical highly excited eigenstates. On the other hand, the average magic of the thermal Scrooge ensemble follows a smooth, continuous profile. Note that in order to show the sharp dips in the eigenstate magic, we do not bin the energy density, but rather plot the magic of each eigenstate individually. Excluding the scar states, and binning the data by energy density, the eigenstate magic closely follows the Scrooge ensemble magic, similar to the results shown for the MFIM in the main text.

\section{Robustness of magic under local unitary circuits}
\label{sm:robustness}

Let $\ket{\psi}$ be an equilibrium pure state of a geometrically $k$-local $N$-qubit Hamiltonian $H$ at effective inverse temperature $\beta$. We write $H = \sum_{Q \in \mathcal{P}_N} a_Q Q$, where each Pauli operator $Q$ with $a_Q \neq 0$ acts on at most $k$ nearby qubits. For concreteness, $\ket{\psi}$ may be an energy eigenstate of $H$ or a state obtained by evolving a generic initial product state to late times under $H$. We consider a $d$-dimensional system and a depth-$t$ unitary circuit $U$ composed of nearest-neighbor gates. We now argue that the magic of $\ket{\psi}$ is robust under the action of $U$.

If $\ket{\psi}$ is an energy eigenstate of $H$, then $U\ket{\psi}$ is an energy eigenstate of the transformed Hamiltonian $H' = UHU^\dag$ with the same energy and effective inverse temperature $\beta$. For $\beta \neq 0$ and $\alpha \geq 2$, the result in the main text gives
\begin{equation}
    \frac{\widetilde{M}_\alpha(\psi)}{N} = \frac{1}{\alpha - 1}\sparens{1 - c_\alpha(H) \beta^{2\alpha} + \bigO{\beta^{2\alpha + 1}}} \,,
\end{equation}
where
\begin{equation}
    c_\alpha(H) = \frac{1}{N\ln 2}\sum_{Q \in \mathcal{P}_N} |a_Q|^{2\alpha}.
\end{equation}
Assuming that $U\ket{\psi}$ is also an equilibrium pure state of $H'$, we have
\begin{equation}
    \frac{\widetilde{M}_\alpha(U\psi)}{N} = \frac{1}{\alpha - 1}\sparens{1 - c_\alpha(H') \beta^{2\alpha} + \bigO{\beta^{2\alpha + 1}}}.
\end{equation}
Write $H' = \sum_{Q \in \mathcal{P}_N} b_Q Q$, where $b_Q = \sum_{P \in \mathcal{P}_N} O_{QP}a_P$ and the orthogonal matrix $O$ represents conjugation by $U$ in the Pauli basis. A large value of $c_\alpha(H')$ means that the coefficients $b_Q$ are concentrated on relatively few Pauli operators. Suppose that, for each output Pauli operator $Q$, at most $L_t$ Pauli terms $P$ with $a_P \neq 0$ contribute to $b_Q$. Equivalently, the restriction of $O$ to the nonzero coefficients of $H$ has at most $L_t$ nonzero entries in each row. The $\ell_2$ norm of this restriction is at most one, while its $\ell_\infty$ norm is at most $\sqrt{L_t}$. Interpolation therefore gives
\begin{equation}
    \norm{Oa}_{2\alpha} \leq L_t^{\frac{1}{2} - \frac{1}{2\alpha}}\norm{a}_{2\alpha} \,,
\end{equation}
where $\norm{a}_p$ denotes the $\ell_p$ norm of the coefficient vector $a = (a_Q)_{Q \in \mathcal{P}_N}$. It follows that
\begin{equation}
    c_\alpha(H') \leq L_t^{\alpha - 1}c_\alpha(H).
\end{equation}
It remains to estimate $L_t$. If $O_{QP} \neq 0$, then $Q$ appears in the Pauli expansion of $UPU^\dag$, so the support of $Q$ lies within the forward light cone of the local operator $P$. Equivalently, by writing $O_{QP}$ as the coefficient of $P$ in $U^\dag Q U$, the input operator $P$ must lie within the backward light cone of $Q$. Therefore, all Pauli operators $P$ that can contribute to a fixed $Q$ must lie within a region of radius $\bigO{t}$ around the support of $Q$.  For the geometrically local Hamiltonian $H$, the number of terms in this region is proportional to its volume, giving $L_t = \bigO{t^d}$. Hence
\begin{equation}
    c_\alpha(H') \leq \bigO{t^{d(\alpha - 1)}}c_\alpha(H).
\end{equation}
The fSRE density can therefore decrease by at most
\begin{equation}
    \frac{\widetilde{M}_\alpha(\psi) - \widetilde{M}_\alpha(U\psi)}{N}
    \leq \bigO{\frac{c_\alpha(H)}{\alpha - 1}\beta^{2\alpha}t^{d(\alpha - 1)}}
    + \bigO{\beta^{2\alpha + 1}}.
\end{equation}
Thus, within the high-temperature expansion, the volume-law magic is robust whenever $\beta^{2\alpha}t^{d(\alpha - 1)} \ll 1$. In particular, any fixed-depth geometrically local circuit produces only a perturbatively small change in the magic density at sufficiently high temperature.

\end{document}